\documentclass[pre,aps,superscriptaddress,preprint]{revtex4-1}

\usepackage{amssymb,amsmath}
\usepackage{graphicx}
\usepackage{nicefrac}
\usepackage{color}
\usepackage{upgreek}

\usepackage{dcolumn}
\usepackage{bm}
\newcommand\p[2]{\frac{\partial #1}{\partial #2}}
\newcommand\pp[3]{\frac{\partial ^{#3} #1}{\partial #2 ^ {#3}}}
\newcommand\ti[1]{\widetilde{#1}}

\begin{document}

\title
{Full self-similar solutions of the subsonic radiative heat equations}
\author
{Tomer Shussman}
\email{tomer.shussman@mail.tau.ac.il}
\affiliation{Raymond and Beverly Sackler School of Physics \& Astronomy, Tel Aviv University, Tel Aviv 69978, ISRAEL}
\affiliation{Department of Plasma Physics, Soreq Nuclear Research Center, Yavne 81800, ISRAEL}
\author
{Shay I. Heizler}
\email{highzlers@walla.co.il}
\affiliation{Department of Physics, Bar-Ilan University, Ramat-Gan, IL52900 ISRAEL}
\affiliation{Department of Physics, Nuclear Research Center-Negev, P.O. Box 9001, Beer Sheva 84190, ISRAEL}

\begin{abstract}

We study the phenomenon of diffusive radiative heat waves (Marshak waves) under general boundary conditions. In particular, we derive full
analytic solutions for the subsonic case, that include both the ablation and the shock wave regions.
Previous works in this regime, based on the work of [R. Pakula and R. Sigel, Phys. Fluids. 443, {\bf 28}, 232 (1985)], 
present self-similar solutions for the ablation region alone, since in general, the shock region and the ablation region are not self-similar together. Analytic results for
both regions were obtained only for the specific case in which the ratio between the ablation front velocity and the shock velocity is constant.
In this work, we derive a full analytic solution
for the whole problem in general boundary conditions. Our solution is composed of two different self-similar solutions, one for each region, that are patched at the heat front.
The ablative region of the heat wave is solved in a manner similar to previous works. Then, the pressure at the front,
which is derived from the ablative region solution, is taken as a boundary condition to the shock region, while the other boundary is described by Hugoniot relations. 
The solution is compared to full numerical simulations in several representative cases. The numerical and analytic results are found to agree within $1\%$ in the ablation region,
and within $2-5\%$ in the shock region. This model allows better prediction of the physical behavior of radiation induced shock waves,
and can be applied for high energy density physics experiments.

\end{abstract}

\maketitle
 
\section{Introduction}

Radiation heat waves play important roles in many high energy density physics (HEDP) phenomena. In particular, they have major importance in inertial confinement fusion
(ICF) and in astrophysical and laboratory plasma~\cite{zeldovich,lindl,rosen,rosenScale,rosenScale2,mihalas}. In these experiments, laser beams deliver energy to the interior of a high-Z hohlraum that is
converted into x-rays. Re-emission and further absorption of the x-rays in the cavity walls helps achieving a thermal source which acts as the drive for the experiments. The radiation is absorbed and
contained within the cavity in a form of a radiative heat wave propagating through the hohlraum walls. It is therefore important to understand this phenomena,
as a key to interpreting the experiments and the numerical simulations.

\begin{figure}
\centering{
(a)
\includegraphics*[width=7cm]{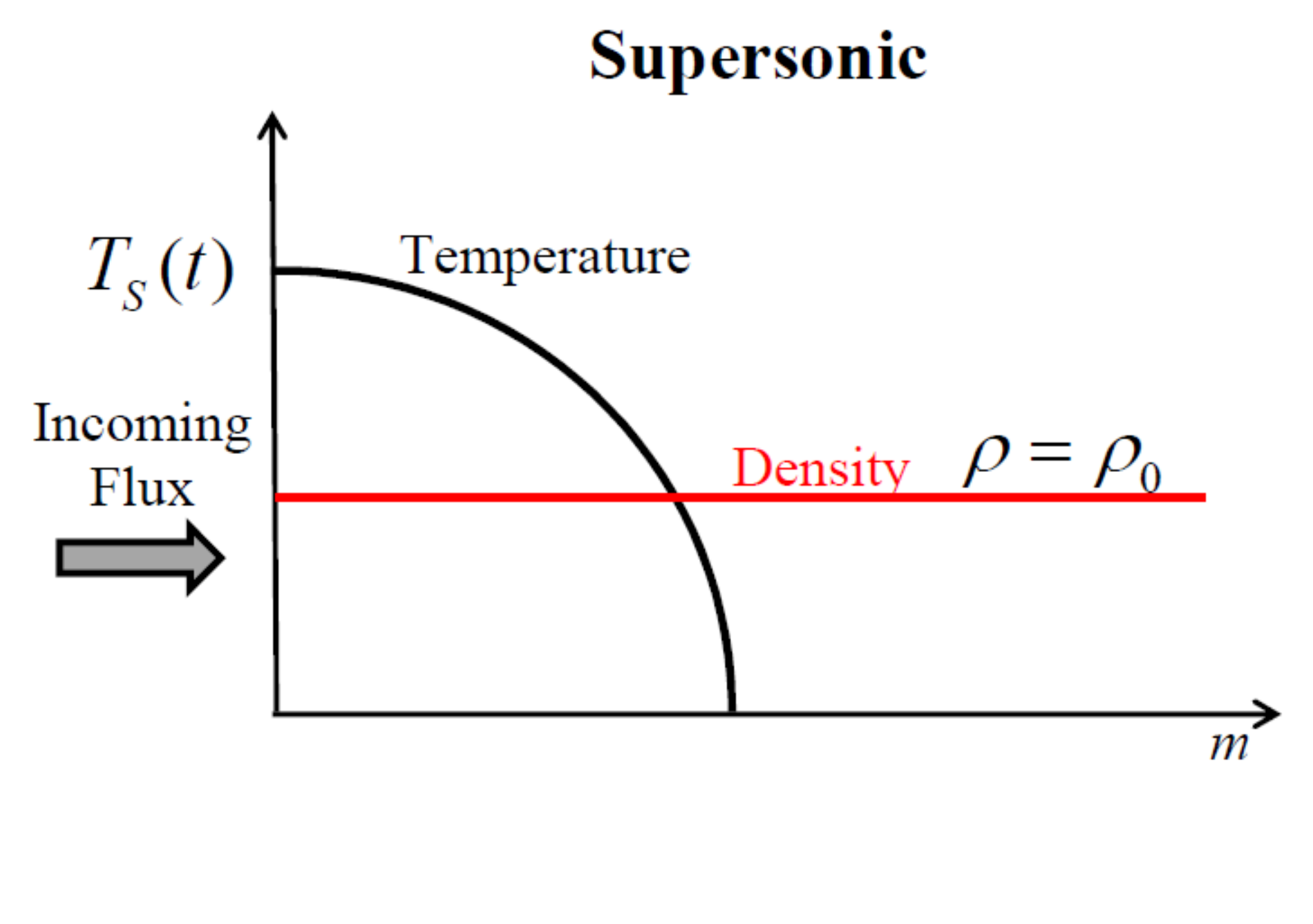}
(b)
\includegraphics*[width=8cm]{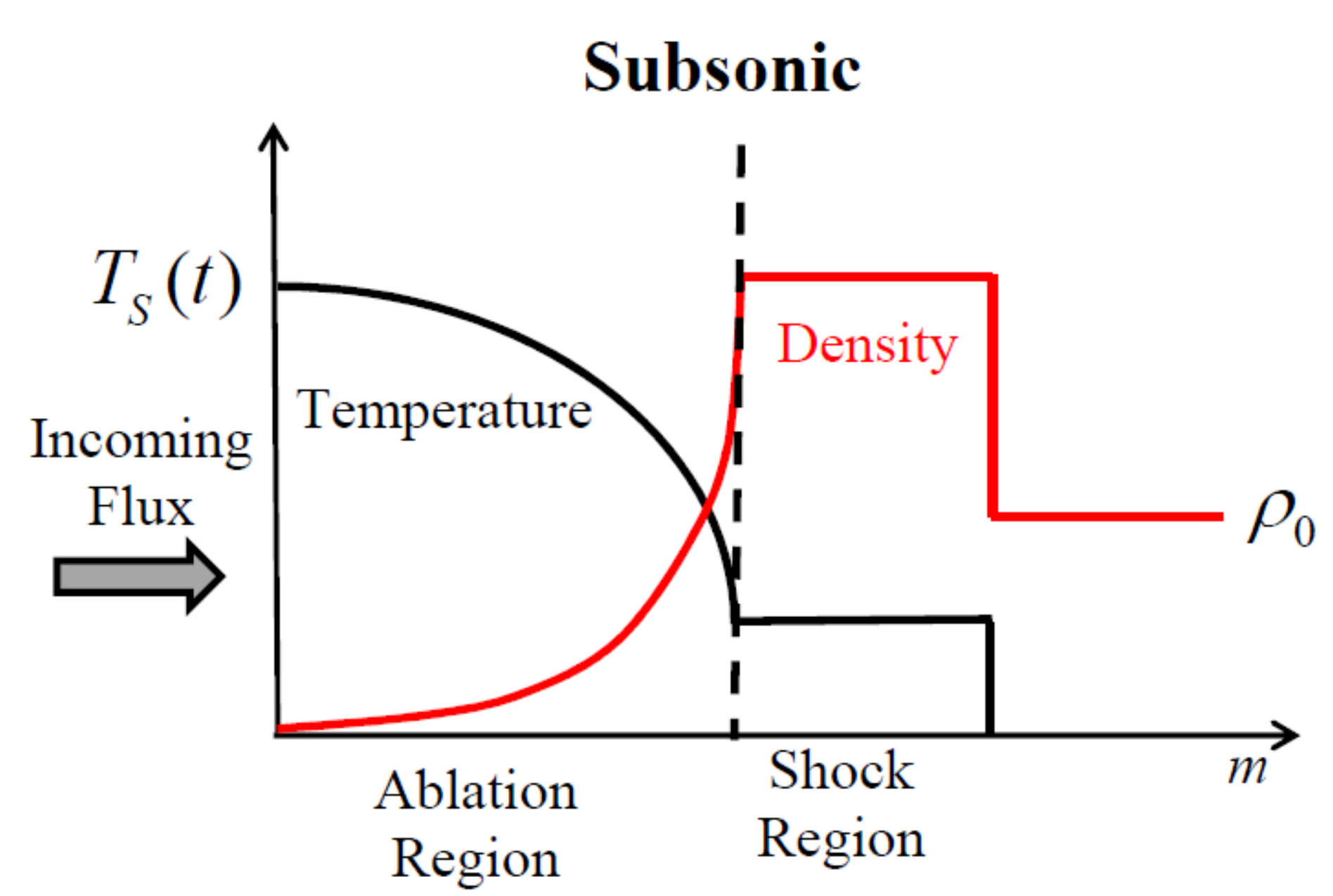}
}
\caption{(Color online) (a)  A schematic diagram of a supersonic radiative heat wave. The heat wave propagates faster than the speed of sound and thus, hydrodynamic motion is negligible and the
density is unchanged. The boundary condition on the surface is of a time dependent temperature.
$m$ is the Lagrangian coordinate. (b) A schematic diagram of a subsonic wave. Two separate regions exist: the ablation region, in which the density is low and the heat flux is dominant (left), and the shock region, in which the density is
high and the heat flux is negligible (right).}
\label{plots_basic}
\end{figure}

The mechanism of the radiative heat waves is as follows: We consider a semi-infinite wall, whose boundary is held at a high temperature. Usually, the temperature and density in this regime are such that the radiation energy is negligible compared to the matter
energy, but the radiation heat flux is the dominant energy transport mechanism~\cite{zeldovich,pombook}. The hot boundary radiates and heats the rest of the wall via photon transport. In the optically thick limit, a diffusive heat
wave, characterized by a sharp temperature rise, propagates through the wall. If the wave propagates much faster than the speed of sound, hydrodynamic motion is negligible in the problem, and the wave is considered to
be ``supersonic" (Fig. \ref{plots_basic}(a)). If, however, the wave propagates slower than the speed of sound, the high matter pressure causes ablation of matter in the opposing direction of the heat wave.
In addition, the heat wave is overtaken by a shock wave (Fig. \ref{plots_basic}(b)), generated by the ablation pressure (from momentum conservation). 
The nature of the heat wave is temperature and density dependent, and can vary with time, as a supersonic
diffusive front decelerates and becomes subsonic if the boundary temperature doesn't change by much. If the temperature rises fast enough, the diffusive front accelerates 
and becomes more and more supersonic. In this work, we assume that the radiation and matter are in local thermodynamic equilibrium (LTE), which means they are strongly coupled.  

The first description of heat waves was proposed by Marshak~\cite{marshak}, who obtained exact solutions for the radiative flow in the supersonic regime, in which hydrodynamic motion is
negligible. For the subsonic case, in which hydrodynamics cannot be neglected, a full self-similar solution combining the ablation region and the shock region cannot be proposed for the general case,
since the problem is not self-similar altogether.
Pakula and Sigel~\cite{ps,ps2,ger3}, obtained self
similar solutions of the heat wave in the subsonic case of an infinitely dense wall, thus solving the ablation region only. Hammer and Rosen~\cite{hr,rh} proposed new solutions in this region,
based on a perturbation theory. Self-similar solutions of the ablative subsonic regime in specific cases were obtained in many other works~\cite{garnier,dubois,murakami,sanz,smith,saillard}.
In particular, Garnier et. al.~\cite{garnier} proposed a self-similar solution that includes both the ablative and shock regions,
for a specific case that ensures constant density over time (which is in this particular case, self-similar). These solutions are widely
used for obtaining a better understanding of the heat wave phenomenon, evaluating the achieved temperature in ICF experiments~\cite{rosen,rosenScale,rosenScale2,rosenScale3},
or modeling hydrodynamical instabilities via linear perturbation amplification technique~\cite{abeguile}.

However, none of the previous works provides a full treatment of the shock wave and the ablation-shock interface for 
the general case. Only naive approximations, considering a constant ablation pressure were used to describe the shock region~\cite{tlusty,kauf,kauf2,hatchett,saillard}.
These approximations are inaccurate in many
cases, where the ablation pressure varies significantly over time. In this work, we propose a complete solution for both parts of the heat wave. We solve each part separately,
in a self-similar fashion, and find a continuous way to mathematically patch
them. For the ablative region we follow the solution of Ref.~\cite{ps} and expand it, finding a relation between the units of the conserved physical quantity and the temporal behavior
of the boundary condition. Furthermore, we derive explicit expressions for the heat wave front coordinate and the energy contained within the heat wave, for general boundary conditions.
In the shock region, we obtain a self-similar solution assuming a time-dependent pressure boundary condition. Then,
the pressure at the ablation front, as obtained in the ablation region solution, is used as the boundary condition of the shock solution. The full solution is composed of these self-similar solutions. 
The results agree with full numerical simulation of the problem.

This paper proceeds as follows: In Secs. \ref{supersonic} and \ref{subsonic} we solve the supersonic and the subsonic heat wave in the ablation region, using the mechanism of~\cite{ps}.
In Sec. \ref{shock} we present the shock wave solution. In Sec. \ref{full} we patch the solutions, provide numerical simulation results, and compare 
them with the integrated solution. A short discussion is presented in Sec. \ref{discussion}.

\section{supersonic waves}
\label{supersonic}

\subsection{Statement of the problem}
We consider a semi-infinite wall of matter at density $\rho_0$. At time $t_0$ the wall-vacuum interface is brought into contact with a thermal bath whose temperature is $T(t)$.
If hydrodynamic motion is negligible, the radiative heat transport is described by one equation alone~\cite{zeldovich,pombook,heizler}:
\begin{equation} \label{super_basic}
\rho_0\frac{\partial e}{\partial t} = \frac{\partial}{\partial x}\left( \frac{cl_R}{3}\frac{\partial}{\partial x}(aT^4) \right)
\end{equation}
Where $l_R=\nicefrac{1}{\kappa_R\rho}$ is the Rossland mean free path, $e$ is the internal energy per unit mass, and $a\equiv\nicefrac{4\sigma}{c}$ is the radiation
density constant ($\sigma$ is Stefan-Boltzmann constant) and $c$ is the speed of light. We assume the opacity and internal energy of the matter can be expressed in the form of power
laws of the density and temperature, and follow the notation of~\cite{hr}:
\begin{subequations} \label{pwrlaws}
\begin{equation}
\frac{1}{\kappa_R}=gT^\alpha\rho^{-\lambda}
\end{equation}
\begin{equation} \label{pwrlaw_energy}
e=fT^\beta\rho^{-\mu}
\end{equation}
\end{subequations}
Using this assumption, Eq. \ref{super_basic} becomes
\begin{equation}
AT^{\beta-1}\frac{\partial T}{\partial t} = \frac{\partial}{\partial m} \left( T^{\alpha+3}\frac{\partial T}{\partial m} \right)
\end{equation}
when $A=3f\beta \rho_0^{-\mu+\lambda}/16\sigma g$ is a dimensional constant typical to this problem, and $m=\rho_0x$ is the Lagrangian coordinate.
If we further assume that the boundary temperature is given as a power law,
\begin{equation} \label{temperature_pwrlaw}
T(t)=T_0t^{\uptau},
\end{equation} 
then the problem has three typical units, mass [$M$], length [$L$] and time [$\theta$], and is characterized by exactly three parameters: 
\begin{subequations} \label{super_dimensions}
\begin{equation}
[t]=[\theta]
\end{equation}
\begin{equation}
[T_0]=[L]^{\frac{2}{\beta}}[\theta]^{-\frac{2}{\beta}-\uptau}
\label{super_temp_units}
\end{equation}
\begin{equation}
[A]=[M]^{-2}[L]^{\frac{8+2\alpha+2\beta}{\beta}}[\theta]^{\frac{-2+3\beta-8}{\beta}}
\end{equation}
\end{subequations}
This means that a self-similarity of the first kind exists~\cite{zeldovich}, and a construction of any dimensional variable using these parameters is unique.
In this analysis, we distinguish between the two variables ($m$, $t$) of the problem and its three physical units. Although the problem is two-dimensional in the usual manner,
it possesses three different physical units.
We note that the temperature is related to the internal energy through Eq. \ref{pwrlaw_energy}, so the units of the temperature can be defined by
forcing $f\rho_0^{-\mu}$ to be dimensionless. The problem can also be solved using four units and $f$ as an additional dimensional constant.

\subsection{The self-similar equation}
Using the parameters and the dimensions given above, a dimensionless parameter connecting the Lagrangian coordinate (with dimensions of $[M]^{1}[L]^{-2}$) and the temporal coordinate can be constructed:
\begin{equation} \label{super_dim_less}
\xi=mA^{\frac{1}{2}}T_0^{\frac{\beta-\alpha-4}{2}}t^{\frac{\uptau(\beta-\alpha-4)-1}{2}}
\end{equation}
The temperature profile through the wall is therefore given as:
\begin{equation} \label{temp_profile}
T(m,t)=T_0t^\uptau \widetilde{T}(\xi)
\end{equation}
Eq. \ref{super_dim_less} also implies
\begin{subequations} \label{derivatives}
\begin{equation}
\frac{\partial \widetilde{T}}{\partial m} =  \frac{\xi}{m} \frac{\partial \widetilde{T}}{\partial \xi}
\end{equation}
\begin{equation}
\frac{\partial \widetilde{T}}{\partial t} = \frac{w_3 \xi}{t} \frac{\partial \widetilde{T}}{\partial \xi}
\end{equation}
\end{subequations}
Here, $w_3\equiv[\uptau(\beta-\alpha-4)-1]/2$ is the temporal power of $\xi$. Substituting these relations into Eq. \ref{super_basic} yields a dimensionless Ordinary differential equation (ODE)
\begin{equation} \label{super_ss}
\ti{T}^{\beta-\alpha-4} \left(\uptau \ti{T}+w_3\xi \p{\ti{T}}{\xi}\right)=(\alpha+3)\ti{T}^{-1}\left(\p{\ti{T}}{\xi}\right)^2+\pp{\ti{T}}{\xi}{2}
\end{equation}
The boundary conditions are $\ti{T}(\xi_F)=0, \p{\ti{T}}{\xi}|_{\xi\to\xi_F}\to-\infty$ and the parameter $\xi_F$ is
determined uniquely by the normalization condition $\ti{T}(0)=1$. The total energy is given by Eq. \ref{pwrlaws}:
\begin{align}
\label{super_energy}
& E(t)=\int_{0}^{m_F} e(\rho,T)dm=f\rho_0^{-\mu}T_0^{\beta}t^{\beta t} \frac{m_F}{\xi_F} \int_{0}^{\xi_F} \ti{T}^{\beta}(\xi)d\xi= \\
& f\rho_0^{-\mu}A^{-\frac{1}{2}}T_0^{\frac{4+\alpha+\beta}{2}}t^{\frac{\uptau(4+\alpha+\beta)}{2}+\frac{1}{2}}\int_{0}^{\xi_F} \ti{T}^{\beta}(\xi)d\xi \nonumber
\end{align}
The albedo can be determined from the relation between the absorbed energy and the emitted flux
\begin{equation}
\frac{1-\alpha}{\alpha}=\frac{\dot{E}(t)}{\sigma T_0t^{4\uptau}}
\end{equation}

\subsection{Boundary condition and conserved quantities}
For a given material, with known $\alpha,\beta,\lambda,\mu$ the value of $\uptau$ fully determines the self-similar solution of Eq. \ref{super_ss}. The parameter $\uptau$ is strongly related to the units of the conserved quantity of
the problem. If $K$ as a conserved quantity of dimensions
\begin{equation}
[K] = [M]^{\lambda_1}[L]^{\lambda_2}[\theta]^{\lambda_3},
\end{equation}
the self-similarity of the problem assures us that a dimensionless constant relates $K$ to the other dimensional parameters, $A$, $T_0$, and $t$. The conservation of $K$ means that this relation is independent of $t$:
\begin{equation}
\bar{\xi}=KA^{w_1}T_0^{w_2}
\end{equation}
Using Eq. \ref{super_dimensions} we deduce:
\begin{subequations}
\begin{equation}
w_1=\frac{\lambda_1}{2}
\end{equation}
\begin{equation}
w_2=-\frac{\beta\lambda_2}{2}-\frac{\alpha+\beta+4}{2}\lambda_1
\end{equation}
\begin{equation}
\uptau=-\frac{2}{\beta}+\frac{\frac{\lambda_1}{\beta}(2\alpha+8-3\beta)-2\lambda_3}{\beta\lambda_2+(\alpha+\beta+4)\lambda_1}
\end{equation}
\end{subequations}
Similarly to $\xi_F$, the dimensionless constant $\bar{\xi}$ is determined by the normalization of the temperature profile and $K$. For example, the case of constant net heat flux through the boundary, $S_0$ is characterized by dimensions
\begin{equation}
[S_0] = [M]^{1}[L]^{0}[\theta]^{-3}
\end{equation}
This yields $\uptau=\frac{1}{4+\alpha+\beta}$ and
\begin{equation}
S_0=\bar{\xi} A^{-\frac{1}{2}} T_0^{\frac{4+\alpha+\beta}{2}}
\end{equation}
The physical meaning of $S_0$ is that the total energy obeys $E(t)=S_0t$. Substituting this in Eq. \ref{super_energy} 
yields the constant value of $\bar{\xi}$
\begin{equation}
\bar{\xi}=f\rho^{-\mu}\int_{0}^{\xi_F} \ti{T}^{\beta}(\xi)d\xi
\end{equation}

\subsection{Solution of the equation}
For solving the equations, one must find $\xi_F$ for which $\ti{T}(0)=1$. This can be done using a shooting method,
or using the self-similar coordinate relation (see Appendix \ref{nispach}). 
Once solved, the numeric value of $\xi_F$ and the self-similar profile can be used to obtain quantitative expressions for the heat front Lagrangian coordinate and the total absorbed energy per unit area, for given surface temperature and time. The expressions are of the form:

\begin{subequations}
\label{super_T0_general}
\begin{equation}
m_F=m_0\rho^{\frac{\mu-\lambda}{2}}T_0^{\frac{4-\beta+\alpha}{2}}t^{\frac{\uptau(4-\beta+\alpha)+1}{2}}
\end{equation}
\begin{equation}
E=e_0\rho_0^{-\frac{\mu+\lambda}{2}}T_0^{\frac{4+\alpha+\beta}{2}}t^{\frac{\uptau(4+\alpha+\beta)}{2}+\frac{1}{2}}
\end{equation}
\end{subequations}

As an example, we take a medium of Au, and use the values shown in~\cite{hr} for the opacity and Equation of state (EOS) of the material.  
The values are specified in Table \ref{table:pwr_law_opac_eos}. Solving for the case of constant surface temperature yields the results:

\begin{subequations}
\label{super_T0_quant}
\begin{equation}
m_F=11.53\cdot 10^{-4}\rho_0^{-0.03}T_0^{1.95}t^{0.5}  \left[\mathrm{\frac{g}{cm^2}}\right]
\end{equation}
\begin{equation}
E=0.29\rho_0^{-0.17}T_0^{3.55}t^{0.5}  \left[\mathrm{\frac{hJ}{mm^2}}\right]
\end{equation}
\end{subequations}

Where $T_0$ is measured in HeV and $t$ is measured in nsec. For the case of constant boundary absorbed heat flux, which is important for hohlraum energy balance
analysis~\cite{rosen,kauf}, the temperature obeys $T(t)=T_0t^{0.1408}$ and we obtain the results:

\begin{subequations}
\label{super_S0_quant}
\begin{equation}
m_F=8.79\cdot 10^{-4}\rho_0^{-0.03}T_0^{1.95}t^{0.775}  \left[\mathrm{\frac{g}{cm^2}}\right]
\end{equation}
\begin{equation}
E=0.21\rho_0^{-0.17}T_0^{3.55}t  \left[\mathrm{\frac{hJ}{mm^2}}\right]
\end{equation}
\end{subequations}

\begin{figure}
\centering{
(a)
\includegraphics*[width=7.4cm]{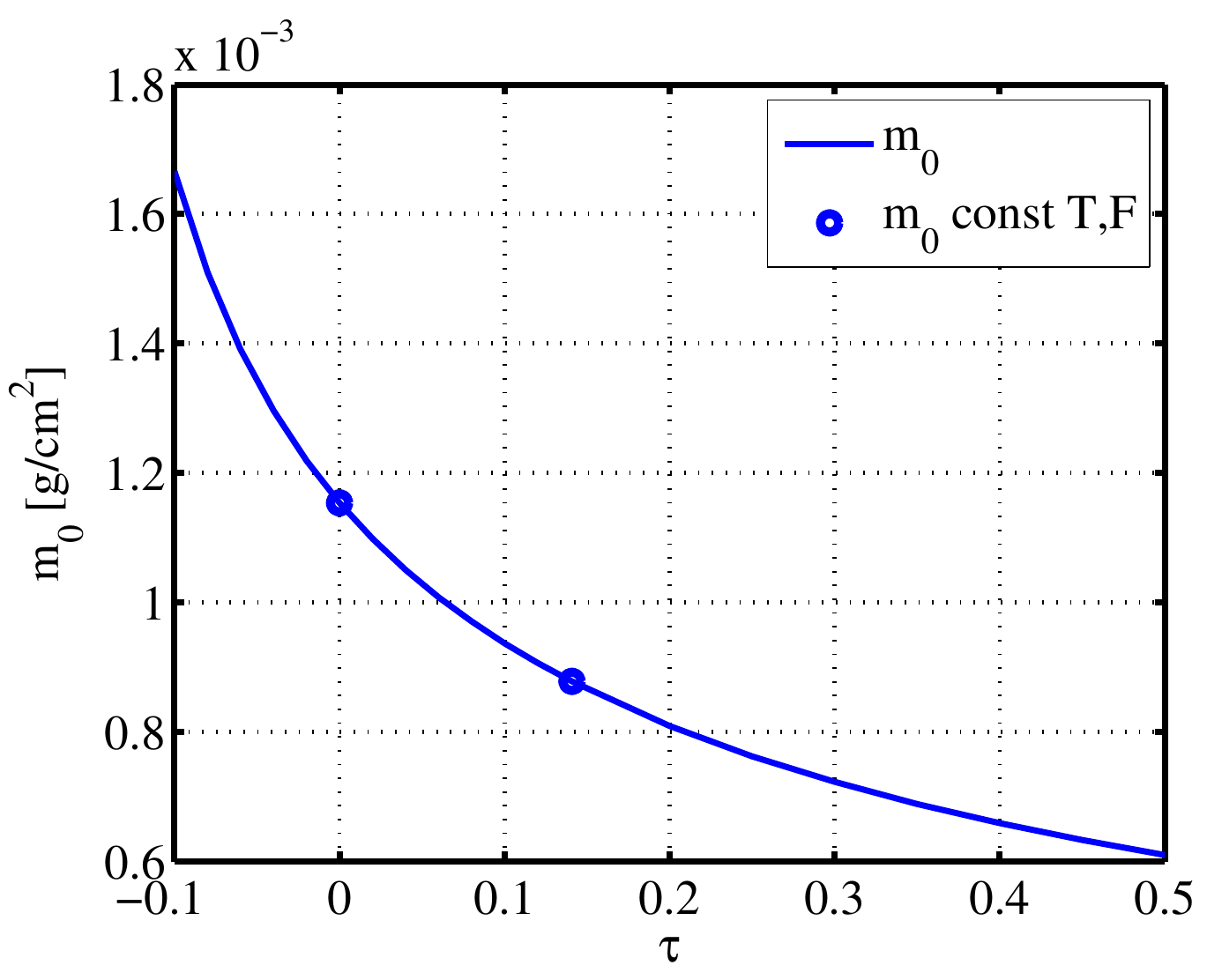}
(b)
\includegraphics*[width=7.4cm]{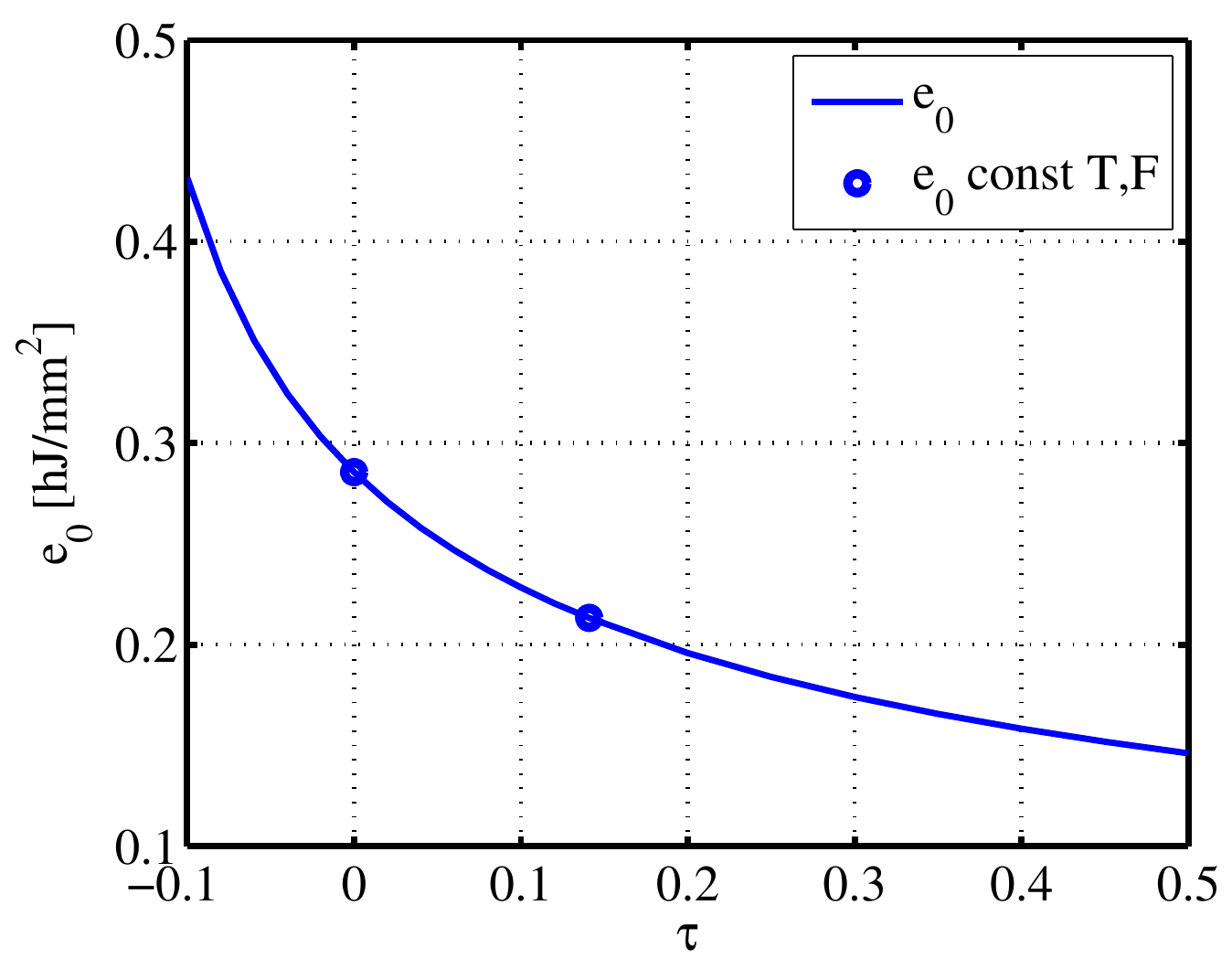}
}
\caption{(Color online) (a) the constant of the Lagrangian coordinate given by Eq. \ref{super_T0_general}(a).
In circles, are marked the two special cases of constant boundary temperature (Eq. \ref{super_T0_quant}(a)) and constant absorbed heat flux (Eq. \ref{super_S0_quant}(a)).
(b) The constant of the energy per unit surface given by Eq. \ref{super_T0_general}(b). In circles, are marked the two special cases of constant boundary temperature (Eq. \ref{super_T0_quant}(b)) and constant absorbed heat flux (Eq. \ref{super_S0_quant}(b)).}
\label{plots_super}
\end{figure}

Quantitative expressions for the general boundary conditions are given in Fig. \ref{plots_super}. 
Numerical simulations (which will be presented in Sec. \ref{shock}) yield the same expressions to within $1\%$ accuracy.

\begin{table}[!htb]
  \centering
  \caption{\bf Power law fits for the opacity and EOS of Au in temperatures $1-3\mathrm{HeV}$~\cite{hr}}
  \label{table:pwr_law_opac_eos} 
  \begin{tabular}{|c|c|} \hline 
    \multicolumn{1}{|c|}{{\bf Physical Quantity}} &
    \multicolumn{1}{c|}{{\bf Numerical Value}} \\ \hline
    \ $f$ & $3.4$ [MJ/g]  \\ \hline
    \ $\beta$ & $1.6$  \\ \hline
    \ $\mu$ & $0.14$  \\ \hline
    \ $g$ & $1/7200$ $[\mathrm{g/cm^2}]$  \\ \hline
    \ $\alpha$ & $1.5$  \\ \hline
    \ $\lambda$ & $0.2$  \\ \hline
    \ $r\equiv(\gamma-1)$ & $0.25$  \\ \hline
  \end{tabular}
\end{table}

\section{subsonic Waves}
\label{subsonic}

\subsection{Statement of the problem}
In the subsonic case, the speed of sound exceeds the heat front velocity, and the hydrodynamic motion is not negligible. The ablated matter density is much lower than the initial bulk density,
while at the heat wave front the matter density is high enough to halt the thermal heat conduction. At this point, the high ablation pressure drives a shock through the bulk.
In order to calculate the behavior of the system, one must solve the full radiative hydrodynamic equations (conservation of mass, momentum, and energy):
\begin{subequations}
\label{sub_basic}
\begin{equation}
\label{mass}
\p{V}{t}-\p{u}{m}=0
\end{equation}
\begin{equation}
\label{momentum}
\p{u}{t}-\p{P}{m}=0
\end{equation}
\begin{equation}
\label{sub_energy}
\p{e}{t}+P\p{V}{t}=\p{}{m}\left(\frac{c}{3\kappa_R}\p{\left(aT^4\right)}{m}\right)
\end{equation}
\end{subequations}
Here, $V\equiv\nicefrac{1}{\rho}$ is the specific volume, $u$ is the matter velocity, $P$ is the pressure, and $m(x,t)=\int_0^x \rho(x',t)dx'$ is the Lagrangian coordinate.
We assume the heat capacity and Rossland mean opacity follow Eq. \ref{pwrlaws}, and the EOS is well described by an ideal gas:
\begin{equation}
P(\rho,T)=r\rho e(\rho,T)\equiv(\gamma-1)\rho e(\rho,T)
\label{def_r}
\end{equation}
From these relations we solve for the temperature:
\begin{equation}
T=\left(\frac{PV^{1-\mu}}{rf}\right)^{\frac{1}{\beta}}
\label{def_T}
\end{equation}
Substituting Eqs. \ref{def_r} and \ref{def_T} in Eq. \ref{sub_energy} yields:
\begin{equation}
\frac{1}{r}\p{PV}{t}+P\p{V}{t}=B\p{}{m}\left(V^{\lambda}\p{}{m} \left( PV^{1-\mu}\right)^{\frac{4+\alpha}{\beta}}\right)
\label{energy2}
\end{equation}
Here, $B$ is a dimensional parameter which is defined as:
\begin{equation}
B=\frac{16\sigma}{3(4+\alpha)}g( rf)^{\frac{-4+\alpha}{\beta}}  
\label{def_B}
\end{equation}
The problem again has three dimensions, and is characterized by three dimensional parameters:
\begin{subequations} \label{sub_dimensions}
\begin{equation}
[t]=[\theta]
\end{equation}
\begin{equation}
[T_0]=[L]^{\frac{2-3\mu}{\beta}}[M]^{\frac{\mu}{\beta}}[\theta]^{-\frac{2}{\beta}-\uptau}
\end{equation}
\begin{equation}
[B]=[M]^{\frac{2\beta+\lambda\beta-\mu(4+\alpha)}{\beta}}[L]^{\frac{3\mu(4+\alpha)-2\alpha-2\beta-3\lambda\beta-8}{\beta}}[\theta]^{\frac{8-3\beta+2\alpha}{\beta}}
\end{equation}
\end{subequations}
In this case we define $f$ to be dimensionless (instead of $f\rho_0^{-\mu}$ in the supersonic case), and thus the units of $T_0$ are now different from Eq. \ref{super_temp_units}.
We note that there exists another dimensional parameter in the problem, the initial density $\rho_0$. This parameter can be neglected in the highly subsonic regime, 
and in the ablation region only, due to the fact that the matter density in the ablation region is much lower than $\rho_0$. Therefore, it can be assumed that the density goes from
$0$ at the rear surface and approaches infinity in the ablation front~\cite{ps,hr}.

Neglecting the initial density of the problem, prevents us from solving the shock region and the ablation region altogether. 
In fact, the full problem, which includes both the ablation and shock is not self-similar, since the shock region depends on the initial density of the matter, via Hugoniot relations,
while the ablation region depends on the dimensional parameter $B$ due to the heat flux. Therefore, solving Eqs. \ref{sub_basic} while neglecting the initial density yields a solution 
for the ablation region alone, in which the temperature at the ablation front is zero (instead of the shock temperature), and the front density approaches infinity.
The same discussion can be applied for the flow velocity. For solving the ablation region, we must neglect the flow velocity at the heat front, and assume it approaches $0$

We will now solve the ablation region, following a method similar to the one used in~\cite{ps,hr,garnier}.

\subsection{The self-similar equations}

Every dimensional variable can be parameterized as a power law of the dimensional parameters:
\begin{equation}
X=\ti{X}B^{w_{X_1}}T_0^{w_{X_2}}t^{w_{X_3}}
\label{def_X}
\end{equation}
In addition, the self-similar coordinate is parameterized as following:
\begin{equation}
\xi=mB^{w_{1}}T_0^{w_{2}}t^{w_{3}}
\label{def_xi}
\end{equation}
The powers $w_{X_i}$ and $w_i$ are deduced from equations \ref{sub_dimensions}. Specifically, the Lagrangian coordinate, with units of $[M]/[L]^2$ is given:
\begin{equation} \label{sub_dim_less}
\xi=m\left( B^{\mu-2}T_0^{2\beta-2\alpha-8-\beta\lambda+(4+\alpha)\mu}t^{-2-2(4+\alpha-\beta)\uptau+\mu(3+(4+\alpha)\uptau)-\lambda(2+\beta\uptau)} \right)^{\frac{1}{4+2\lambda-4\mu}}
\end{equation}
The power law dependence of the rest of the dimensional variables is specified in Table \ref{table:pwr_laws}.

\begin{table}[!htb]
  \centering
  \caption{\bf Power law dependence of the physical quantities in the subsonic ablative heat solution}
  \label{table:pwr_laws} 
  \begin{tabular}{|c|c|} \hline 
    \multicolumn{1}{|c|}{{\bf Physical variable}} &
    \multicolumn{1}{c|}{{\bf Power-law dependency}} \\ \hline
    \ Lagrangian &  $m=\xi\left(B^{2-\mu}T_0^{8+2\alpha-2\beta+\beta\lambda-(4+\alpha)\mu} \right.$ \\
    Coordinate & $t^{2+2(4+\alpha-\beta\uptau-\mu(3+(4+\alpha)\uptau)+\lambda(2+\beta\uptau)}\Big)^
    {\frac{1}{4+2\lambda-4\mu}}$  \\ \hline
    \ Temperature &  $T(m,t)=T_0t^{\uptau}\ti{T}$  \\ \hline
    \ Velocity &  $u(m,t)=\ti{u}\left(B^{-\mu}T_0^{\beta(2+\lambda)-(4+\alpha)\mu}t^{\mu+\beta(2+\lambda)\uptau-(4+\alpha)\mu\uptau}\right)^
    {\frac{1}{4+2\lambda-4\mu}}$  \\ \hline
    \ Pressure &  $P(m,t)=\ti{P}\left(B^{1-\mu}T_0^{4+\alpha+\beta\lambda-(4+\alpha)\mu}t^{-1+\mu+(4+\alpha+\beta\lambda)\uptau-(4+\alpha)\mu\uptau}\right)^
    {\frac{1}{2+\lambda-2\mu}}$  \\ \hline
    \ Specific Volume &  $V(m,t)=\ti{V}\left(B^{-1}T_0^{-4-\alpha+2\beta}t^{1-(4+\alpha-2\beta)\uptau}\right)^
    {\frac{1}{2+\lambda-2\mu}}$  \\ \hline
    \ Heat Flux &  $S(m,t)=\ti{S}\left(B^{2-3\mu}T_0^{8+2\alpha+2\beta+3\lambda\beta-3(4+\alpha)\mu} \right.$ \\ 
    & $t^{-2+3\mu+(2(4+\alpha+\beta)+3\beta\lambda)\uptau-3(4+\alpha)\mu\uptau}\Big)^
    {\frac{1}{4+2\lambda-4\mu}}$  \\ \hline
    \ Energy &  $E(m,t)=\ti{E}\left(B^{2-3\mu}T_0^{8+2\alpha+2\beta+3\lambda\beta-3(4+\alpha)\mu}\right.$ \\
    & $t^{2+2\lambda-\mu+(2(4+\alpha+\beta)+3\beta\lambda)\uptau-3(4+\alpha)\mu\uptau}\Big)^
    {\frac{1}{4+2\lambda-4\mu}}$  \\ \hline
  \end{tabular}
\end{table}

Using Table \ref{table:pwr_laws} and the equivalent of Relations \ref{derivatives} for the subsonic case, a set of dimensionless ODEs can be obtained:

\begin{subequations} \label{sub_ss}
\begin{equation}
\left(w_{V3}+w_3\xi\p{}{\xi}\right)\ti{V}-\p{\ti{u}}{\xi}=0
\end{equation}
\begin{equation}
\left(w_{u3}+w_3\xi\p{}{\xi}\right)\ti{u}+\p{\ti{P}}{\xi}=0
\end{equation}
\begin{align}
& \frac{1}{r}\left[(w_{V3}+w_{P3})\ti{P}\ti{V}+w_3\xi\left(\ti{V}\p{\ti{P}}{\xi}+\ti{P}\p{\ti{V}}{\xi}\right)\right]+\ti{P}\left(w_{V3}+w_3\xi\p{}{\xi}\right)\ti{V}= \\
& \frac{4+\alpha}{\beta} \left\{ \lambda\ti{V}^{\lambda-1}\p{\ti{V}}{\xi}\left(\ti{P}\ti{V}^{1-\mu}\right)^{\frac{4+\alpha-\beta}{\beta}}\left[\ti{P}(1-\mu)\ti{V}^{-\mu}\p{\ti{V}}{\xi}+
\ti{V}^{1-\mu}\p{\ti{P}}{\xi}\right]+\right. \nonumber \\
& \frac{4+\alpha-\beta}{\beta}\ti{V}^{\lambda}\left(\ti{P}\ti{V}^{1-\mu}\right)^{\frac{4+\alpha-2\beta}{\beta}}\left[\ti{P}(1-\mu)\ti{V}^{-\mu}\p{\ti{V}}{\xi}+
\ti{V}^{1-\mu}\p{\ti{P}}{\xi}\right]^2 + \nonumber \\
& \ti{V}^{\lambda}\left(\ti{P}\ti{V}^{1-\mu}\right)^{\frac{4+\alpha-\beta}{\beta}}\left[2(1-\mu)\ti{V}^{-\mu}\p{\ti{V}}{\xi}\p{\ti{P}}{\xi}-
\mu(1-\mu)\ti{P}\ti{V}^{-\mu-1}\left(\p{\ti{V}}{\xi}\right)^2+\right.  \nonumber  \\
&  \left.\left.(1-\mu)\ti{P}\ti{V}^{-\mu}\pp{\ti{V}}{\xi}{2}+\ti{V}^{1-\mu}\pp{\ti{P}}{\xi}{2}\right]\right\} \nonumber
\end{align}
\end{subequations}

The boundary conditions are $\ti{V}(\xi_F)=0$, $\p{\ti{V}}{\xi}|_{\xi_F}=0$, $\ti{P}(0)=0$, 
$\p{\ti{P}}{\xi}|_{\xi=0}=0$, $\ti{u}(\xi_F)=0$, and the free parameter $\xi_F$ is determined from the normalization
condition of the temperature, in a manner similar to the supersonic solution. We note that since the temperature appears
in Eq. \ref{sub_ss} only via the relation specified in Eq. \ref{def_T}, the normalized solution should now satisfy
$\ti{T}(0)^{\beta}=\ti{P}(0)\ti{V}(0)^{1-\mu}=1$. Then, the absorbed energy is given by:

\begin{equation}
\label{sub_energy_quantity}
E(t)=\int_{0}^{m_F} \frac{P(m,t)V(m,t)}{r}+\frac{u^2(m,t)}{2}dm=B^{w_{E_1}}T_{0}^{w_{E_2}}t^{w_{E_3}} \int_{0}^{\xi_F} \frac{\ti{P}(\xi)\ti{V}(\xi)}{r}+\frac{\ti{u}^2(\xi)}{2}d\xi
\end{equation}

\subsection{Boundary condition and the conserved quantities}
The self-similarity of the problem assures us that a dimensionless constant relates $K$ to the other dimensional parameters, $B$, $T_0$, and $t$. The conservation of $K$ means that this relation is independent of t:
\begin{equation}
\bar{\xi}=KB^{w_1}T_0^{w_2}
\end{equation}
Using Eq. \ref{sub_dimensions} we deduce:

\begin{subequations}
\begin{equation}
\left[2\beta+\lambda\beta-\mu(4+\alpha)\right]w_1+\mu w_2=\beta\lambda_1
\end{equation}
\begin{equation}
\left[3\mu(4+\alpha)-2\alpha-2\beta-3\lambda\beta-8\right]w_1+(2-3\mu)w_2=\beta\lambda_2
\end{equation}
\begin{equation}
\uptau=-\frac{\left[w_1(8-3\beta+2\alpha)+\beta\lambda_3\right]}{\beta w_2}-\frac{2}{\beta}
\end{equation}
\end{subequations}

Specifically, for the case of constant flux, with units $[S_0]=[M][\theta]^{-3}$, the boundary condition obeys:

\begin{equation}
\uptau=\frac{2-3\mu}{8+2\alpha+2\beta+3\beta\lambda-3(4+\alpha)\mu}
\end{equation}

\subsection{Solution of the equations}

\begin{figure}
\centering{
\includegraphics*[width=7.4cm]{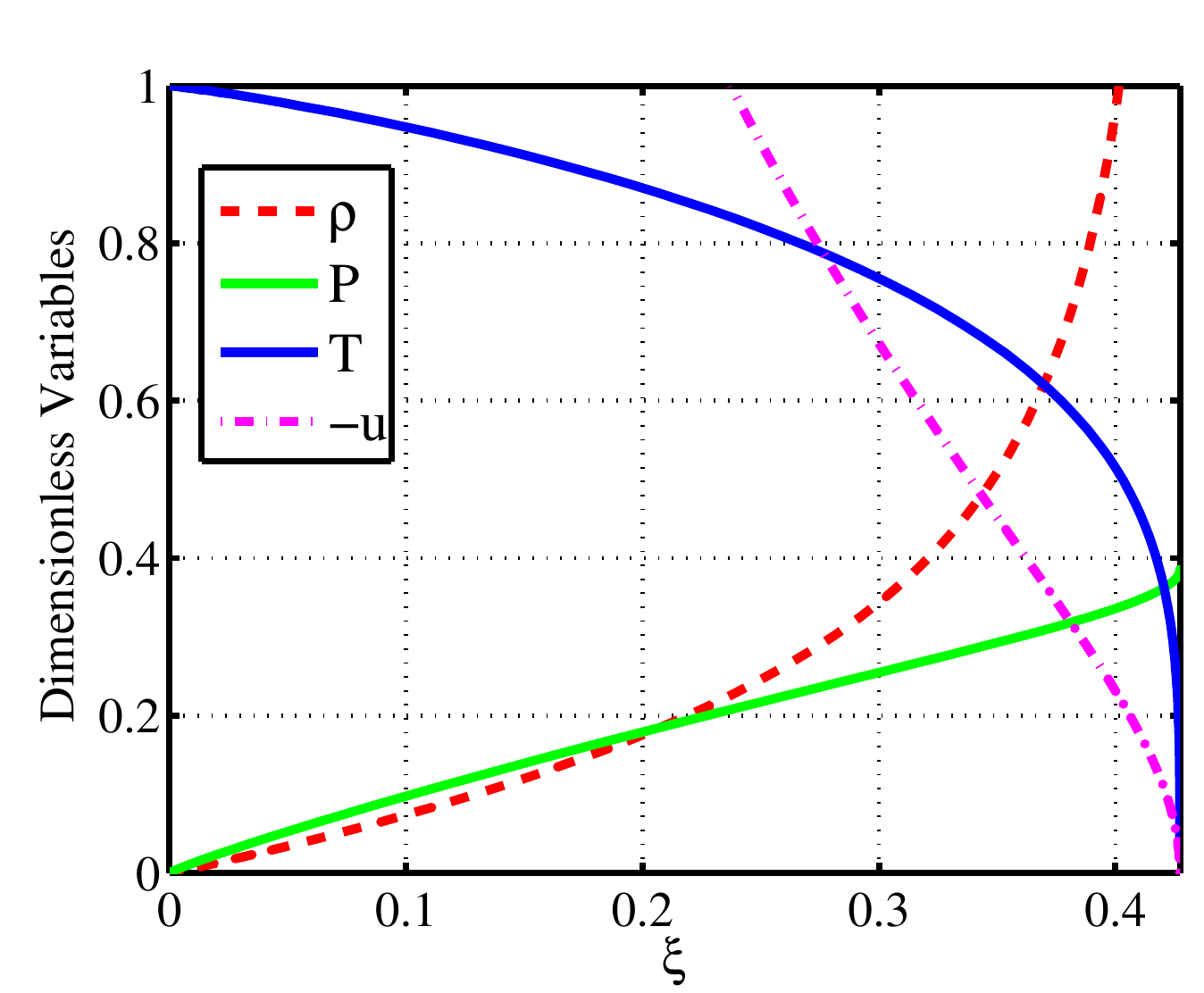}
}
\caption{(Color online) The dimensionless variables, obtained in the numerical integration of Eqs. \ref{sub_ss}, for the case of constant temperature ($\uptau=0$). Since the shock region is neglected, the temperature drops sharply to $0$ at $\xi_F$, and the density diverges. The ablation pressure (e.g. the pressure at the ablation front) has a finite value.}
\label{vars}
\end{figure}

The numerical solution of Eqs. \ref{sub_ss} is obtained by a double shooting method, or by using the self-similar relation (see Appendix \ref{nispach}).
The self-similar profiles for the case of $\uptau=0$, as obtained by the numerical integration of Eqs. \ref{sub_ss} are presented in Fig. \ref{vars}.
Once solving the dimensionless quantities, we can insert them to the self-similar relations given in Table \ref{table:pwr_laws} and obtain quantitative expressions for the heat front
Lagrangian coordinate, total energy, and the ablation pressure at the front. For the case of constant surface temperature:

\begin{subequations}
\label{sub_T0_quant}
\begin{equation}
m_F=10.17\cdot 10^{-4}T_0^{1.91}t^{0.52}  \left[\mathrm{\frac{g}{cm^2}}\right]
\end{equation}
\begin{equation}
E=0.59T_0^{3.35}t^{0.59}  \left[\mathrm{\frac{hJ}{mm^2}}\right]
\end{equation}
\begin{equation}
P_F=2.71T_0^{2.63}t^{-0.45}  \left[\mathrm{Mbar}\right]
\end{equation}
\end{subequations}

For the case of constant boundary net-flux we obtain $T(t)=T_0t^{0.123}$ and:

\begin{subequations}
\label{sub_S0_quant}
\begin{equation}
m_F=8.32\cdot 10^{-4}T_0^{1.91}t^{0.75}  \left[\mathrm{\frac{g}{cm^2}}\right]
\end{equation}
\begin{equation}
E=0.42_0^{3.35}t  \left[\mathrm{\frac{hJ}{mm^2}}\right]
\end{equation}
\begin{equation}
P_F=3.06T_0^{2.63}t^{-0.125}  \left[\mathrm{Mbar}\right]
\end{equation}
\end{subequations}

\begin{figure}
\centering{
(a)
\includegraphics*[width=7.4cm]{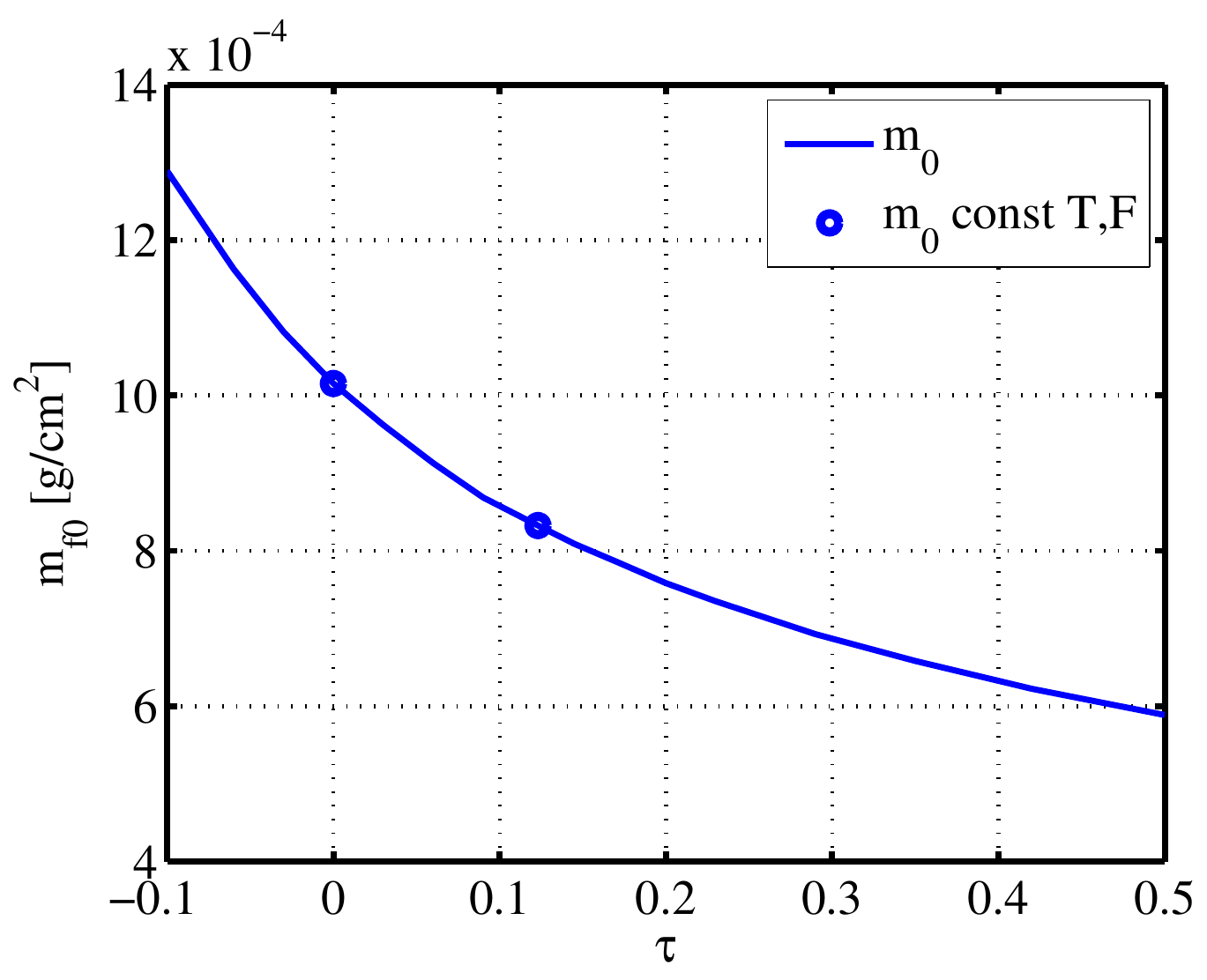}
(b)
\includegraphics*[width=7.4cm]{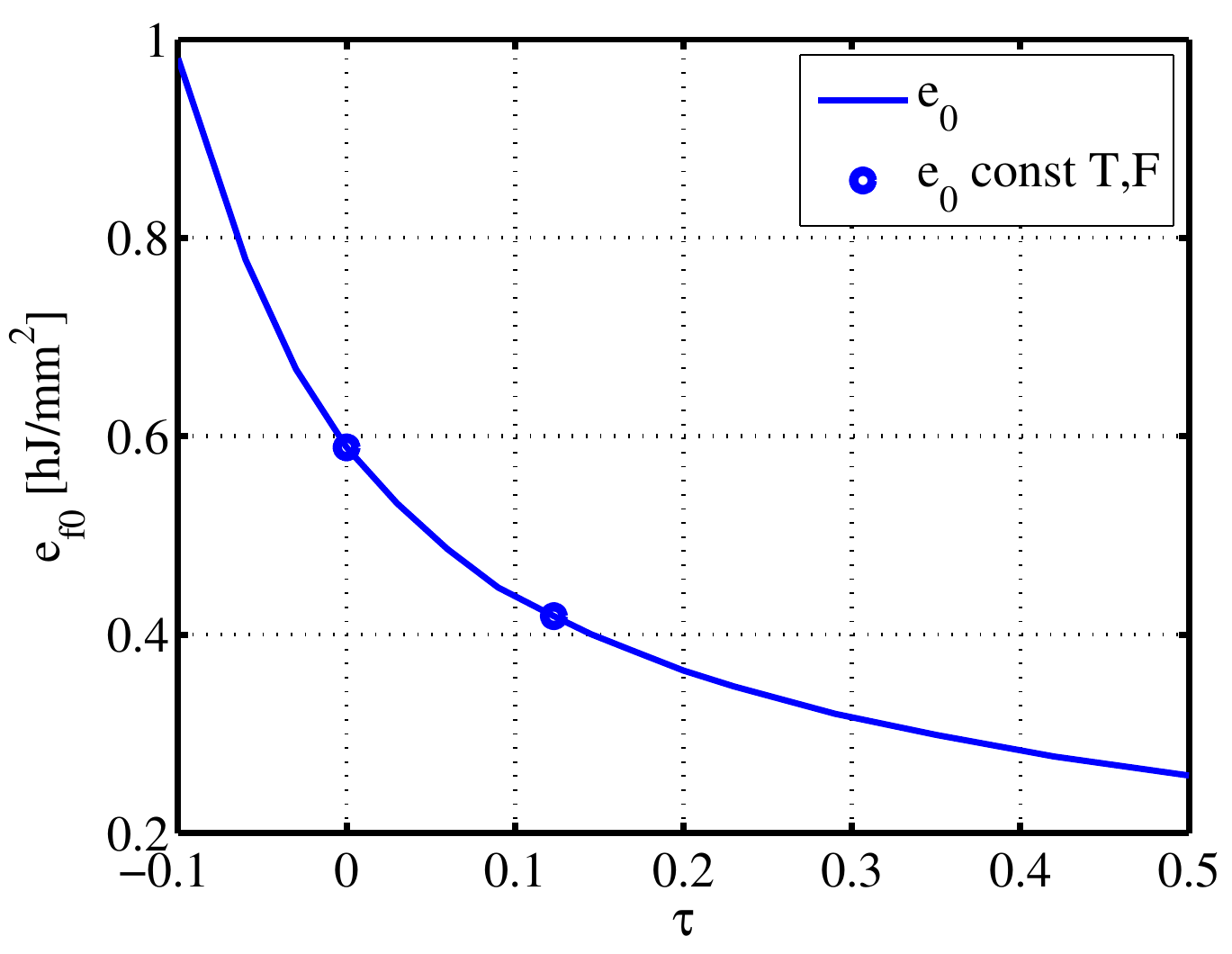}
(c)
\includegraphics*[width=7.4cm]{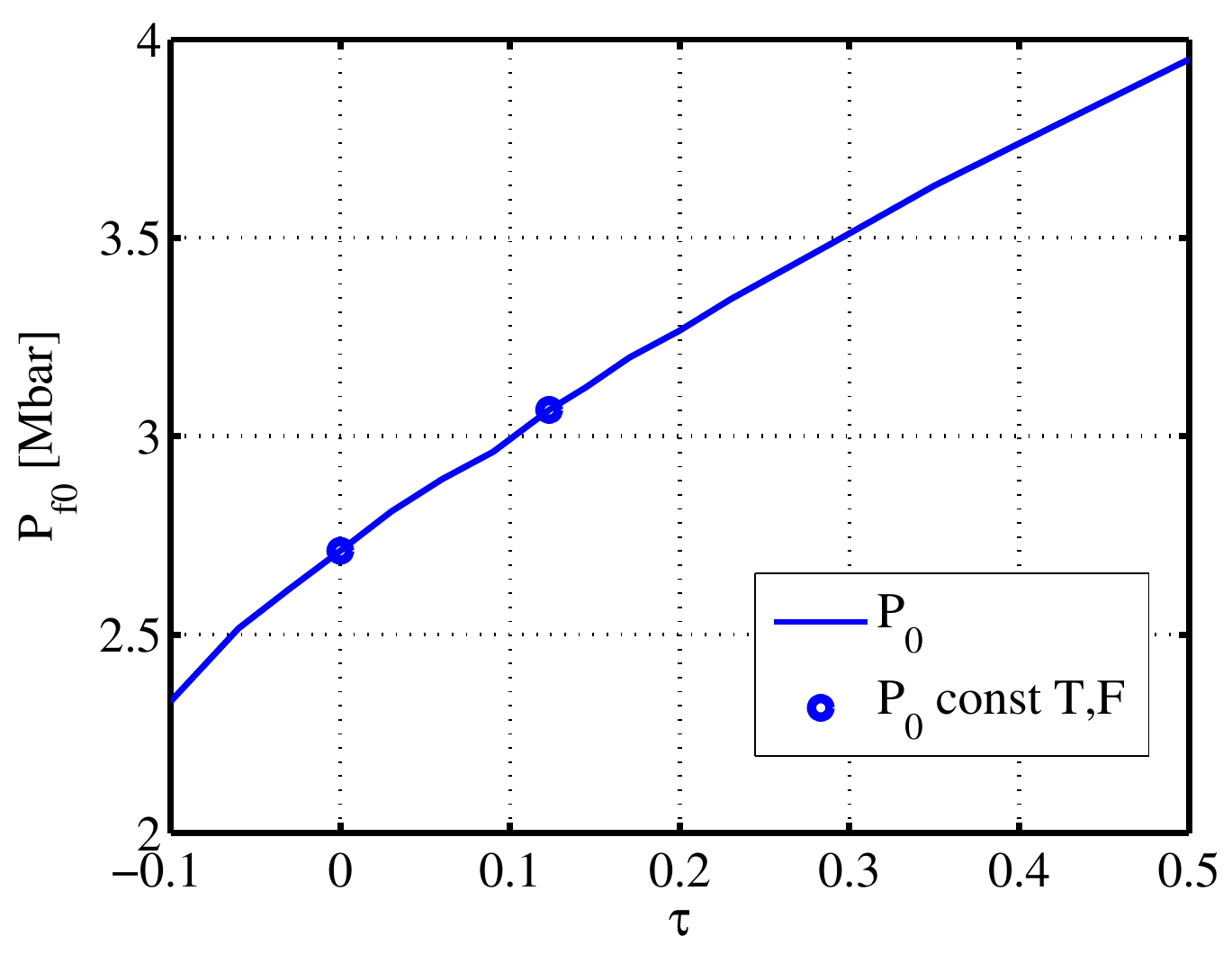}
}
\caption{(Color online) (a) the constant of the Lagrangian coordinate given by Eq. \ref{def_xi}. In circles, are marked the two special cases of constant boundary
temperature (Eq. \ref{sub_T0_quant}(a)) and constant absorbed heat flux (Eq. \ref{sub_S0_quant}(a)). (b) the constant of the energy per unit surface
given by Eq. \ref{def_X}. In circles, are marked the two special cases of constant boundary temperature (Eq. \ref{sub_T0_quant}(b)) and constant
absorbed heat flux (Eq. \ref{sub_S0_quant}(b)). (c) The same for the ablation pressure.}
\label{plots_sub}
\end{figure}

Quantitative expressions for the general boundary conditions are given in Fig. \ref{plots_sub}. Here, again, numerical simulations reproduce the analytic expressions to within $1\%$ accuracy.
In addition, it reproduced both the 
analytic and numeric solutions of the supersonic and subsonic ablative heat waves, which were obtained and checked in previous works~\cite{rosen,rosenScale,rosenScale2,rosenScale3,kauf}.

\section{Shock solution}
\label{shock}

In this section we solve the shock region of the subsonic heat wave. As mentioned earlier, the full problem, composed of the shock region and the ablation region, is not self-similar.
Therefore, we solve the shock region independently, and in the next section we patch both regions and obtain a full solution.

\subsection{Statement of the problem}
Mathematically, the shock region is different from the ablation region by two main parameters. Since the downstream and upstream densities are related via Hugoniot Relations, the initial density $\rho_0$ is a key parameter of the shock region, and it seems as if the shock region is not self-similar at all. However, in the case of a strong shock, as obtained in highly-subsonic heat waves, radiation heat conduction is negligible in the shock region. Therefore we can ignore the dimensional parameter $B$, which is presented in the previous section.

The question remains what is the proper boundary condition to describe the surface between the regions, and three natural contenders are the ablation pressure, the ablation velocity, and the front density. The density is obviously a bad choice since it diverges at the ablation front, and the velocity is assumed to be
zero at the front. In fact, matching the self-similar solution in the ablation region to a self-similar solution in the shock region will only be valid if the flow velocities
in the shocked region are much smaller than the exhaust flow in the ablation region, which is of order the speed of sound of the radiatively-heated material.

Regarding the discussion above, the pressure boundary condition should be satisfying. We use the heat front pressure as a boundary condition, and assume that the shock boundary is well approximated as a surface with power law pressure dependence.

For the shock region in the general case, the set of equations needs to be solved is:
\begin{subequations}
\label{shock_basic}
\begin{equation}
\label{mass_s}
\p{V}{t}-\p{u}{m}=0
\end{equation}
\begin{equation}
\label{momentum_s}
\p{u}{t}-\p{P}{m}=0
\end{equation}
\begin{equation}
\label{energy_s}
\p{e}{t}+P\p{V}{t}=0
\end{equation}
\end{subequations}

We assume the boundary condition:
\begin{equation}
P(t)=P_0t^{\uptau_S}
\end{equation}

In the case of a shock created by a heat wave, this boundary condition is obtained from Table \ref{table:pwr_laws}.
We further assume that the shocked material EOS obeys the ideal gas relation (Eq. \ref{def_r}) where $r$ is not necessarily the same for the ablation region and the shock region.
In the shock front, the boundary conditions obey Hugoniot Relations~\cite{zeldovich}:

\begin{subequations} \label{hugoniot}
\begin{equation}
\frac{D-u}{V_S}=\frac{D}{V_0}
\end{equation}
\begin{equation}
P_S=\frac{Du_S}{V_0}
\end{equation}
\begin{equation}
P_Su_SV_0=D\left(e_S+\frac{u_S^2}{2}\right)
\end{equation}
\end{subequations}

Here, $V_0$ is the unperturbed specific volume, $V_S$, $u_S$, $P_S$, and $e_S$ are the specific volume, matter velocity, pressure, and thermal energy right after the shock, and $D$ is the shock
velocity in the lab frame. The parameters of the problem and their dimensions are:

\begin{subequations} \label{shock_dimensions}
\begin{equation}
[t]=[\theta]
\end{equation}
\begin{equation}
[P_0]=[L]^{-1}[M][\theta]^{-2-\uptau_S}
\end{equation}
\begin{equation}
[V_0]=[L]^{3}[M]^{-1}
\end{equation}
\end{subequations}

Using these parameters, we again deduce self-similar relations of the form:
\begin{equation}
X=\ti{X}P_0^{w_{X_1}}V_0^{w_{X_2}}t^{w_{X_3}}
\label{def_X_shock}
\end{equation}
The power law dependence of the physical variables is given in Table \ref{table:pwr_shock_laws}. Specifically, the self-similar coordinate is:
\begin{equation}
\label{shock_xi}
\xi=m\left(P_0^{-\nicefrac{1}{2}}V_0^{\nicefrac{1}{2}}t^{-1-\frac{\uptau_S}{2}} \right)
\end{equation}

\begin{table}[!htb]
  \centering
  \caption{\bf Power law dependence of the physical quantities in the shock solution}
  \label{table:pwr_shock_laws} 
  \begin{tabular}{|c|c|} \hline 
    \multicolumn{1}{|c|}{{\bf Physical variable}} &
    \multicolumn{1}{c|}{{\bf Power-law dependency}} \\ \hline
    \ Lagrangian Coordinate & $m_S=\xi P_0^{\frac{1}{2}}V_0^{-\frac{1}{2}}t^{1+\frac{\uptau_S}{2}} $  \\ \hline
    \ Velocity &  $u(m,t)=\ti{u}P_0^{\frac{1}{2}}V_0^{\frac{1}{2}}t^{\frac{\uptau_S}{2}}$  \\ \hline
    \ Pressure &  $P(m,t)=\ti{P}P_0t^{\uptau_S}$  \\ \hline
    \ Specific Volume &  $V(m,t)=\ti{V}V_0$  \\ \hline
    \ Energy &  $E(m,t)=\ti{E}P_0^{\frac{3}{2}}V_0^{\frac{1}{2}}t^{1+\frac{3}{2}\uptau_S}$  \\ \hline
  \end{tabular}
\end{table}

\subsection{The Self-Similar equations}

Substituting Eq. \ref{shock_xi} into Eq. \ref{shock_basic}, and using the derivatives $\p{\xi}{t}=-\left(1+\nicefrac{\uptau_S}{2}\right)\frac{\xi}{t}$ and $\p{\xi}{m}=\frac{\xi}{m}$, we obtain the self-similar
set of equations:
\begin{subequations} \label{shock_ss}
\begin{equation}
-\left(1+\frac{\uptau_S}{2}\right)\xi\p{\ti{V}}{\xi}-\p{\ti{u}}{\xi}=0
\end{equation}
\begin{equation}
-\left(1+\frac{\uptau_S}{2}\right)\xi\p{\ti{u}}{\xi}+\frac{\uptau_S}{2}\ti{u}+\p{\ti{P}}{\xi}=0
\end{equation}
\begin{equation}
-\left(1+\frac{\uptau_S}{2}\right)\frac{\xi}{r}\ti{V}\p{\ti{P}}{\xi}+\frac{\ti{V}\ti{P}\uptau_S}{r}-\left(1+\frac{1}{r}\right)\left(1+\frac{\uptau_S}{2}\right)\xi\ti{P}\p{\ti{V}}{\xi}=0
\end{equation}
\end{subequations}

The boundary conditions are $\ti{P}(0)=1$ and the self-similar Hugoniot Relations at $\xi_S$. The shock velocity obeys:

\begin{equation}
D=V_0\dot{m}=P_0^{\nicefrac{1}{2}}V_0^{\nicefrac{1}{2}}t^{\nicefrac{\uptau_S}{2}}\left(1+\frac{\uptau_S}{2}\right)\xi
\label{hugoniot_d}
\end{equation}

Substituting Eq. \ref{hugoniot_d} in Eq. \ref{hugoniot} yields the self-similar Hugoniot relations:
\begin{subequations} \label{hugoniot_ss}
\begin{equation}
\ti{V}_S=1-\frac{\ti{u}_S}{\left(1+\frac{\uptau_S}{2}\right)\xi}
\end{equation}
\begin{equation}
\ti{P}_S=\left(1+\frac{\uptau_S}{2}\right)\xi\ti{u}_S
\end{equation}
\begin{equation}
\left(1+\frac{\uptau_S}{2}\right)\xi_S\left(\frac{\ti{P}_S\ti{V}_S}{r}+\frac{\ti{u}_S^2}{2}\right)=\ti{P}_S\ti{u}_S
\end{equation}
\end{subequations}

Rearranging the equations yields:
\begin{subequations} \label{hugoniot_ss2}
\begin{equation}
\ti{V}_S=\frac{r}{r+2}\left(\ldots=\frac{\gamma-1}{\gamma+1}\right)
\end{equation}
\begin{equation}
\ti{u}_S=\frac{2}{r}\left(1+\frac{\uptau_S}{2}\right)\xi\ti{V}_S
\end{equation}
\begin{equation}
\ti{P}_S=\frac{\ti{u}_S^2}{2}\frac{r}{\ti{V}_S}
\end{equation}
\begin{equation}
\ti{D}=\frac{r+2}{2}\ti{u}_S
\label{shock_velocity}
\end{equation}
\end{subequations}

Once again, the free parameter $\xi_S$ is determined from the normalization
condition at $\xi=0$, but this time the boundary pressure is normalized to match the boundary condition.

\subsection{Solution of the Equations}
The set of Eqs. \ref{shock_ss} is solved using a shooting method, just like in the supersonic case.
After obtaining the self-similar profiles, quantitative expressions for the shock front Lagrangian Coordinate and the bulk velocity are derived, using Table \ref{table:pwr_shock_laws} (The
shock velocity can be calculated using Eq. \ref{shock_velocity}).
For convenience, we choose the same value of $r$ that we used for the ablation region. For a constant boundary pressure we get:

\begin{subequations}
\label{shock_P0_const}
\begin{equation}
m_S=4.66\cdot 10^{-3}P_0^{0.5}t  \left[\mathrm{\frac{g}{cm^2}}\right]
\end{equation}
\begin{equation}
u_S=2.15P_0^{0.5}  \left[\mathrm{\frac{km}{sec}}\right]
\end{equation}
\end{subequations}

The cases of constant surface temperature in the ablation region, and constant absorbed flux are of particular interest for this work.
In the case of constant temperature, the ablation pressure temporal behavior yields (see Eq. \ref{sub_T0_quant}) $w_{P_3}\equiv\uptau_S=-0.45$
(we remind that the notation of $w_{P_i}$ is defined in Eq. \ref{def_X} and calculated in Table \ref{table:pwr_laws}).
Substituting this as a boundary condition for the shock gives the relation: 
\begin{subequations}
\label{shock_T0_const}
\begin{equation}
m_S=7.34\cdot 10^{-3}P_0^{0.5}t^{0.7765}  \left[\mathrm{\frac{g}{cm^2}}\right]
\end{equation}
\begin{equation}
u_S=2.62P_0^{0.5}t^{-0.2235}  \left[\mathrm{\frac{km}{sec}}\right]
\end{equation}
\end{subequations}

In the case of constant absorbed flux, the ablation pressure behaves as $w_{P_3}\equiv\uptau_S=-0.124$, and the numerical expressions are:
\begin{subequations}
\label{shock_S0_const}
\begin{equation}
m_S=5.17\cdot 10^{-3}P_0^{0.5}t^{0.938}  \left[\mathrm{\frac{g}{cm^2}}\right]
\end{equation}
\begin{equation}
u_S=2.23P_0^{0.5}t^{-0.062}  \left[\mathrm{\frac{km}{sec}}\right]
\end{equation}
\end{subequations}

\begin{figure}
\centering{
(a)
\includegraphics*[width=7.4cm]{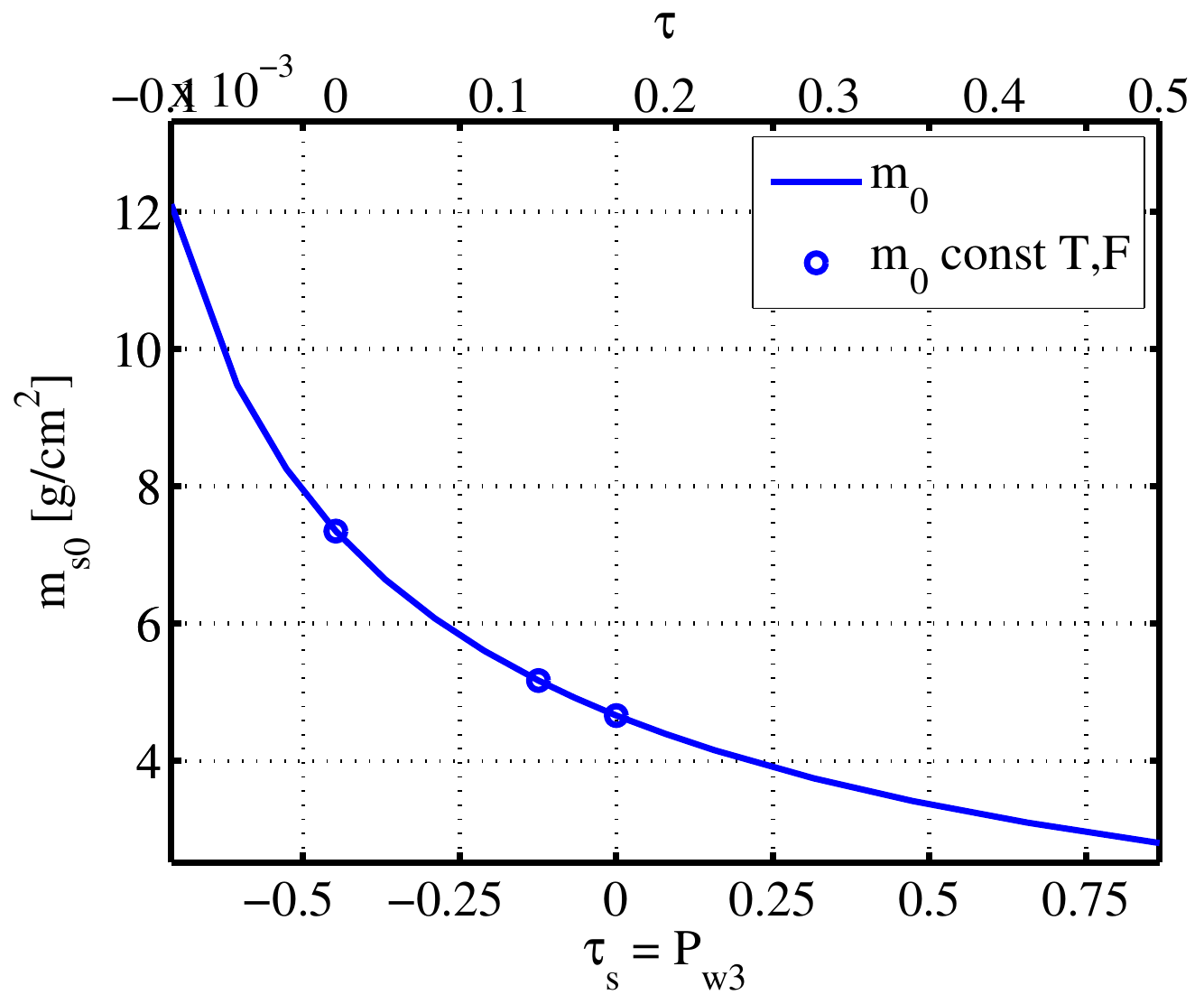}
(b)
\includegraphics*[width=7.4cm]{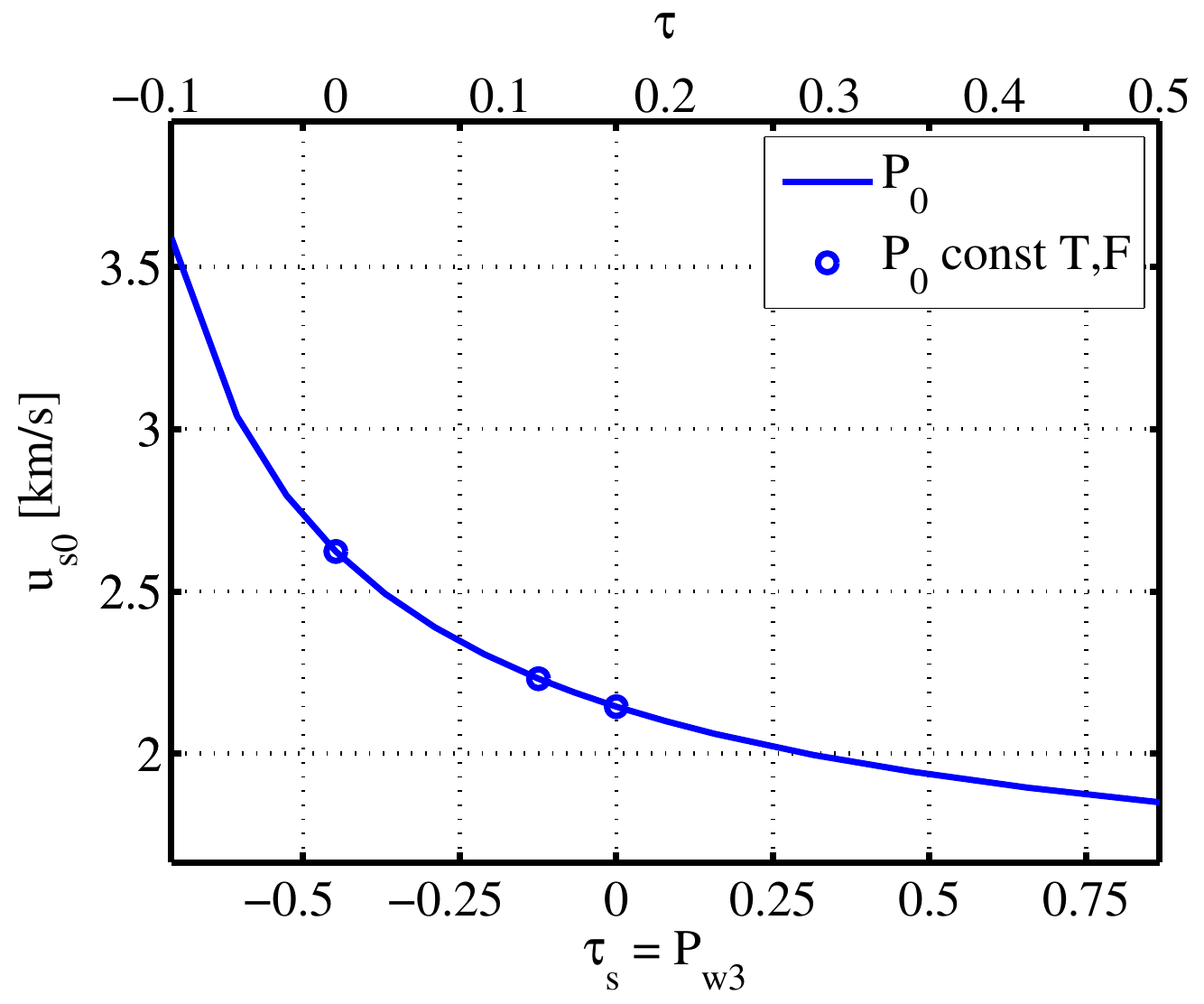}
}
\caption{(Color online) (a) the constant of the Lagrangian coordinate given by Eq. \ref{shock_xi}. In circles, are marked three special cases of constant boundary
temperature (Eq. \ref{shock_T0_const}(a)), constant absorbed heat flux (Eq. \ref{shock_S0_const}(a)) and constant ablation pressure (\ref{shock_P0_const}(a)).
(b) the constant of the matter velocity at the shock front. In circles, are marked three special cases of constant boundary temperature (Eq. \ref{shock_T0_const}(b)), constant
absorbed heat flux (Eq. \ref{shock_S0_const}(b)) and constant ablation pressure (\ref{shock_P0_const}(b)).}
\label{plots_shock}
\end{figure}

Quantitative expressions for the general boundary conditions are given in Fig. \ref{plots_shock}. Using these expressions, one can derive the shock solution for any given
ablation region boundary condition. This can be done by substituting $\uptau_S$ with $w_{P_3}$ and $P_{0,\mathrm{shock}}$ with $P_{0,\mathrm{heat}}T_0^{w_{P_2}}$,
where $P_{0,\mathrm{shock}}$, $P_{0,\mathrm{heat}}$ correspond to the
shock region boundary condition constant, and the ablation pressure constant, respectively. We note that the temperature can be obtained, using Eq. \ref{def_T}. In this work,
for simplicity we assume the same values of $f$, $\beta$, $\mu$ and $r$ for both the ablation regime and the shock regime, although this is usually not the case. 

\section{Full solution and Numerical Simulations}
\label{full}
In this final section, we present a full solution of the subsonic heat equation, decomposed of the two self-similar solutions obtained earlier for the ablation
region (Sec. \ref{subsonic}) and for the shock region (Sec. \ref{shock}). The two solutions are patched together in the following manner:
\begin{figure}
\centering{
\includegraphics*[width=15cm]{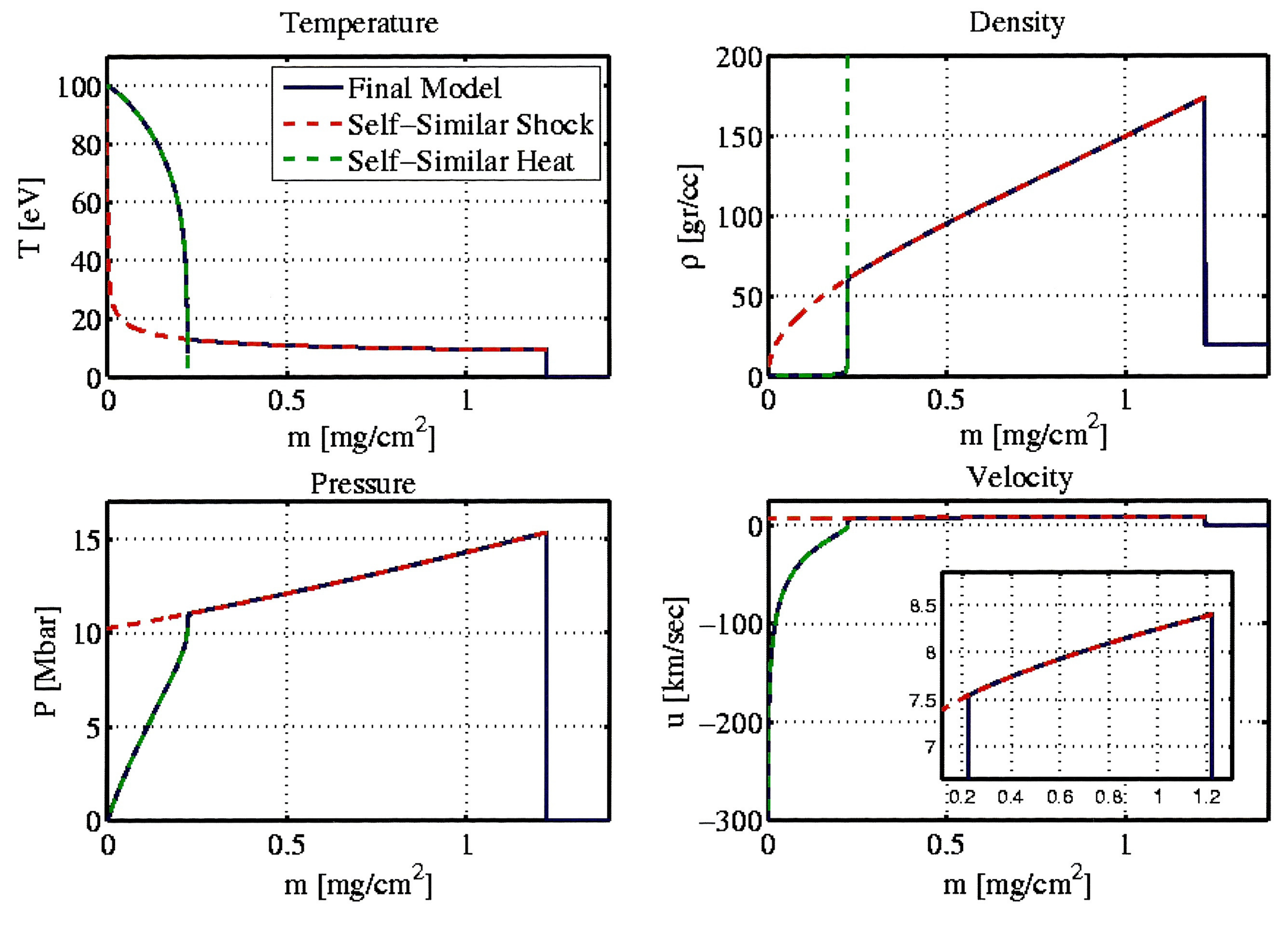}
}
\caption{(Color online) An example of the patching method of the two different regions, in a constant surface temperature problem.
The two self-similar solutions of the two different regions are marked in dashed line, while the full solution
is taken as the ablation solution from $m=0$ to the patching point (when the curves collide), and from there, the shock solution is used. The small box in the velocity
graph is a zoomed look on the shock region.}
\label{Patching}
\end{figure}

\begin{figure}
\centering{
\includegraphics*[width=15cm]{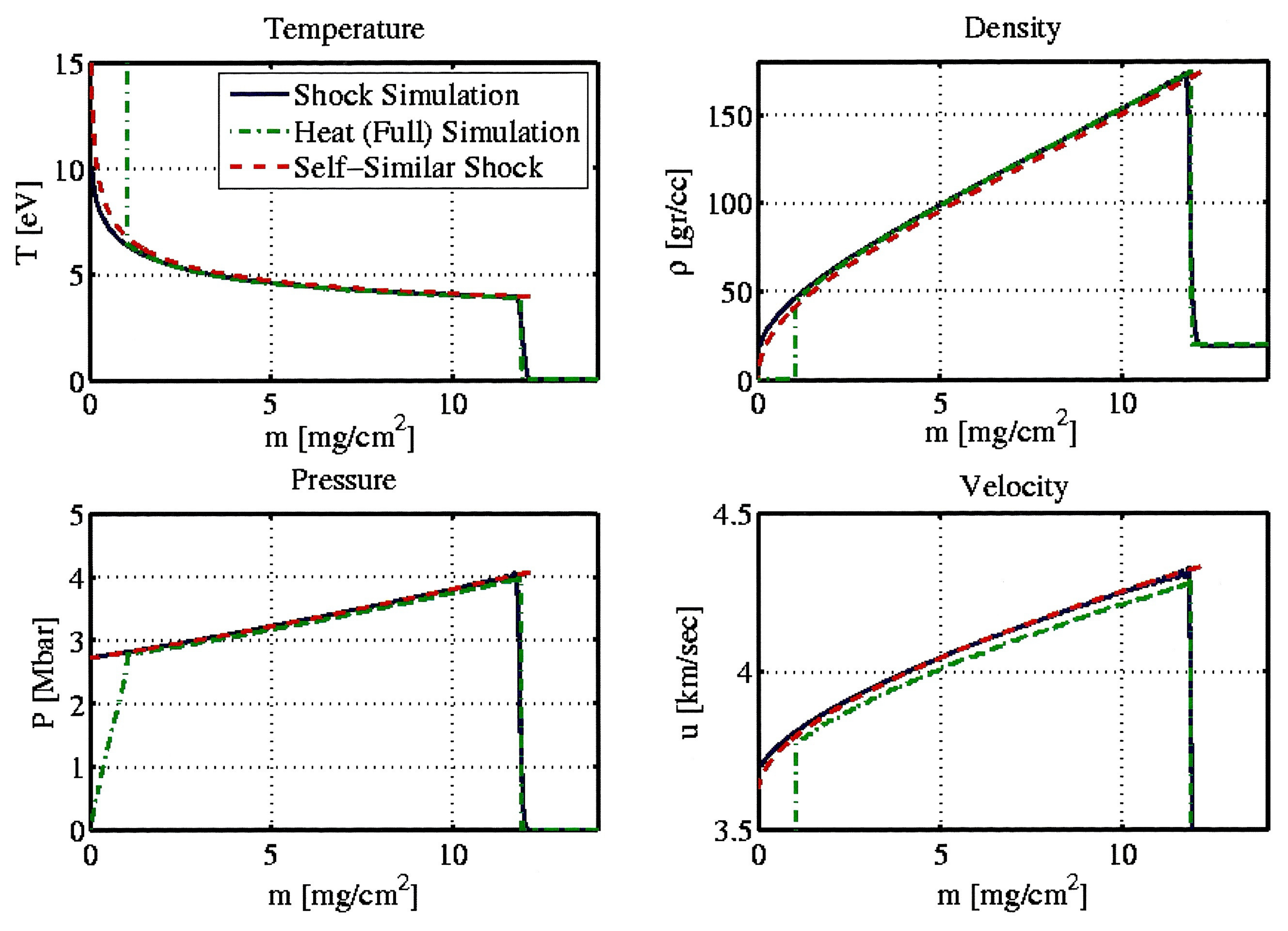}
}
\caption{(Color online) A comparison between the self-similar shock solution and numerical simulations. The dashed (full)
line shows the shock analytic (numerical) solution for a boundary condition of $P=2.71t^{-0.45}\mathrm{Mbar}$.
The dashed-dotted line is yielded from a full heat wave numerical solution in the boundary condition $T=1\mathrm{HeV}$ (this heat wave boundary condition yields an ablation pressure of $P=2.71t^{-0.45}\mathrm{Mbar}$ exactly). The shock numerical solution agrees with the analytic solution
within $1\%$, and the full simulation agrees with the analytic solution to within $3\%$ in the shock region.}
\label{Shock_check}
\end{figure}
\begin{itemize} 
\item The ablation region is solved using a specified boundary condition, given in the form of Eq. \ref{temperature_pwrlaw}.
Self-similar and quantitative profiles for the temperature, pressure, density and velocity are obtained up to the ablation front $m_F$.
\item The ablation pressure is applied as the power law boundary condition of the shock region problem ($P_{0,\mathrm{shock}}=P_{0,\mathrm{heat}}T_0^{w_{P_2}}$,$\uptau_S=w_{P_3}$).
\item The shock region is solved using this boundary condition and Hugoniot relations at the front. Self-similar and quantitative profiles for the temperature, pressure,
density and velocity are obtained from $m=0$ to the shock front $m_S>m_F$.
\item The full solution is given by the ablation region profile from $m=0$ to $m=m_F$, the shock region profile from $m=m_F$ to $m=m_S$ and the undisturbed matter profile from there forward.
\end{itemize} 

An example of the patching method is shown in Fig. \ref{Patching}, for the case of constant temperature boundary condition $T_0=1\mathrm{HeV}$, and at
the time $t=0.05ns$. The solution is shown at an early time so that the full profile can be well seen, as the shock wave propagates ahead of the heat front at later times.

\begin{figure}
\centering{
\includegraphics*[width=15cm]{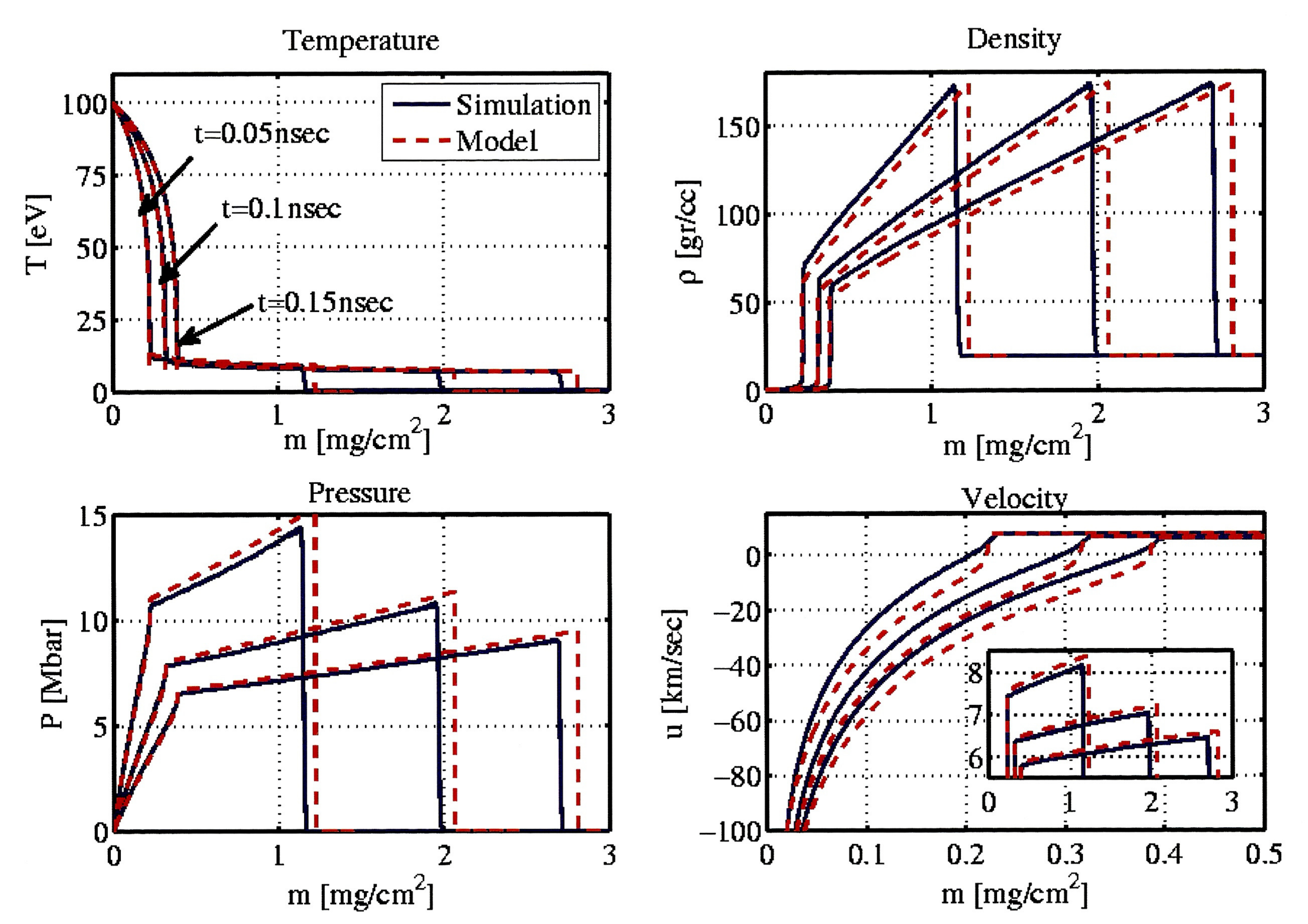}
}
\caption{(Color online) A comparison between the model (dashed lines) and simulations (full lines) under a boundary
condition of constant surface temperature ($\uptau=0$) with $T_0=1\mathrm{HeV}$
in early times. Presented are the profiles of the temperature, density, pressure and matter velocity. The small box in the velocity
graph is a zoomed look on the shock region.}
\label{early_T0}
\end{figure}
\begin{figure}
\centering{
\includegraphics*[width=15cm]{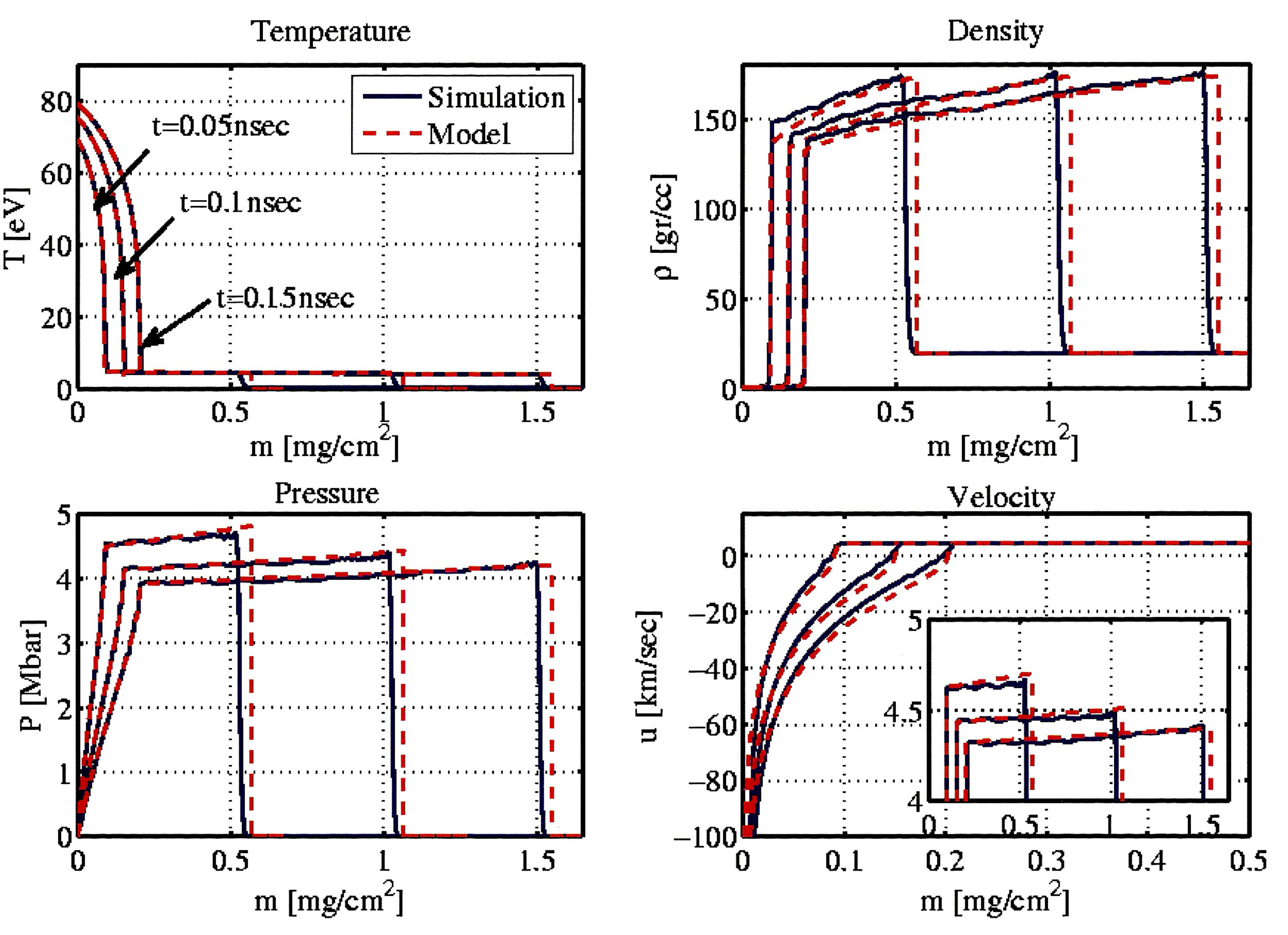}
}
\caption{(Color online) A comparison between the model (dashed lines) and simulations (full lines) under a boundary
condition of constant absorbed flux ($S_0$, $\uptau=0.123$) with $T_0=1\mathrm{HeV}$
in early times. Presented are the profiles of the temperature, density, pressure and matter velocity. The small box in the velocity
graph is a zoomed look on the shock region.}
\label{early_S0}
\end{figure}
\begin{figure}
\centering{
\includegraphics*[width=15cm]{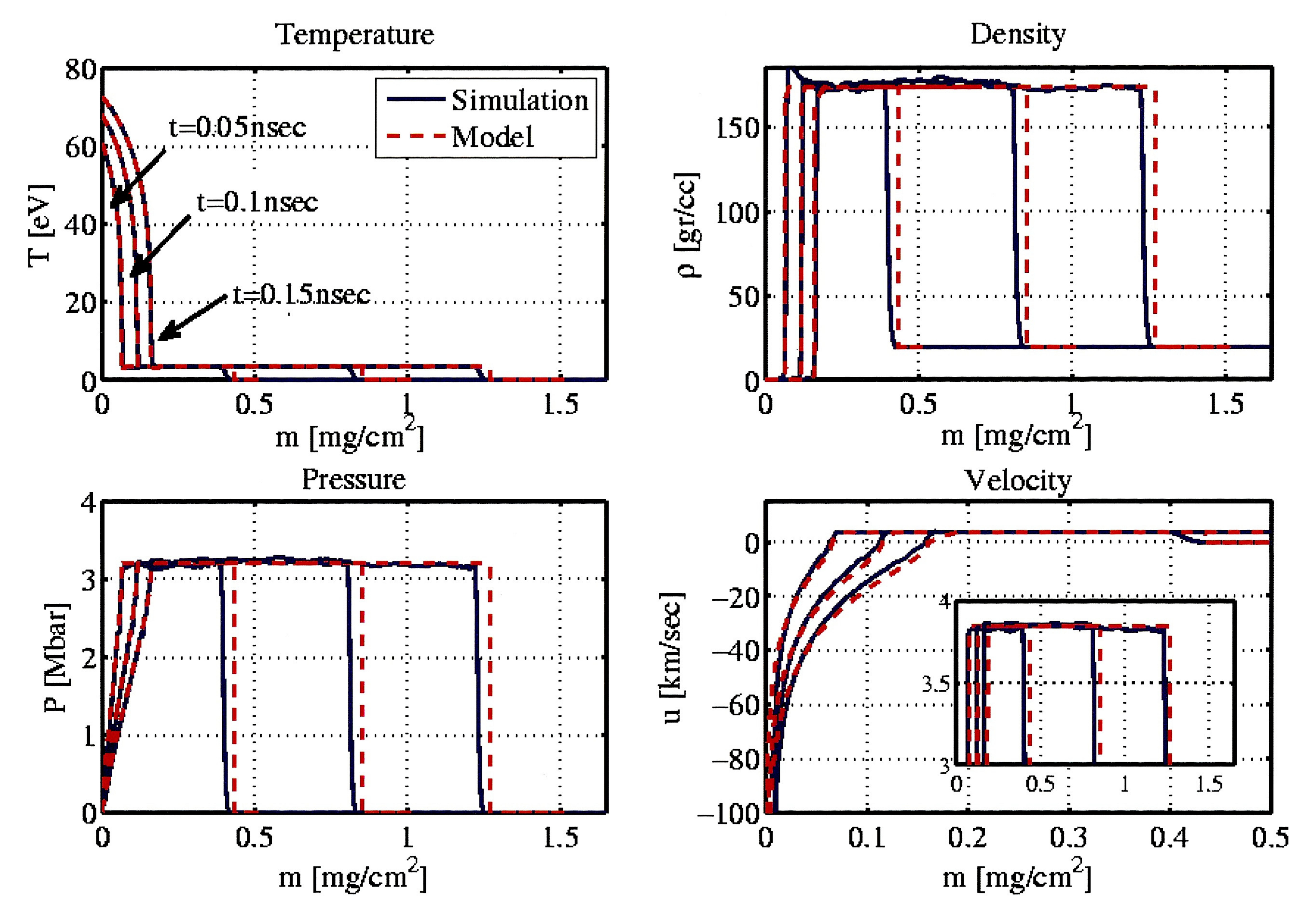}
}
\caption{(Color online) A comparison between the model (dashed lines) and simulations (full lines) under a boundary
condition of constant ablation pressure ($P_0$, $\uptau_S=0$, $\uptau=0.17$) with $T_0=1\mathrm{HeV}$
in early times. Presented are the profiles of the temperature, density, pressure and matter velocity. The small box in the velocity
graph is a zoomed look on the shock region.}
\label{early_P0}
\end{figure}
In order to solve for the self-similar shock region profile, we had to assume that the boundary condition is well described as a surface with power law pressure dependence.
This is obviously not the case, since the ablation front matter velocity is actually not zero, and the density does not diverge, as opposed to the assumptions that led to the analytic ablation region solution. However, comparison of the self-similar solution with the numerical solution of a shock, in the boundary condition of $P_0=2.71$ and
$\uptau_S=-0.447$ (which corresponds to a heat wave of $1$HeV constant temperature),
and with a full numerical simulation of a subsonic heat wave, as shown in Fig. \ref{Shock_check}
shows an agreement of $1\%$ between all three profiles. This is due to the fact that only the boundary of the problem is affected by the velocity
condition, as the rest of the bulk is solely affected by the pressure gradient.

For the numerical simulations we used a standard 1D hydrodynamic Lagrangian code, which includes radiative transport
in 1D LTE diffusion approximation. The code was validated via regular test cases. In addition, it reproduced the 
analytic solutions of the supersonic and subsonic ablative heat waves, which were obtained and checked in previous works~\cite{rosen,rosenScale,rosenScale2,rosenScale3,kauf}.
In the simulations, we used the same power laws for the EOS and the opacity as in the self-similar model.
In the ``full simulation", we set the surface boundary condition to be $T_s(t)=T_0t^{\uptau}$ for different
values of $\uptau$, and checked that the conserved quantities which correspond to each value of $\uptau$ were indeed
conserved.

\begin{figure}
\centering{
\includegraphics*[width=15cm]{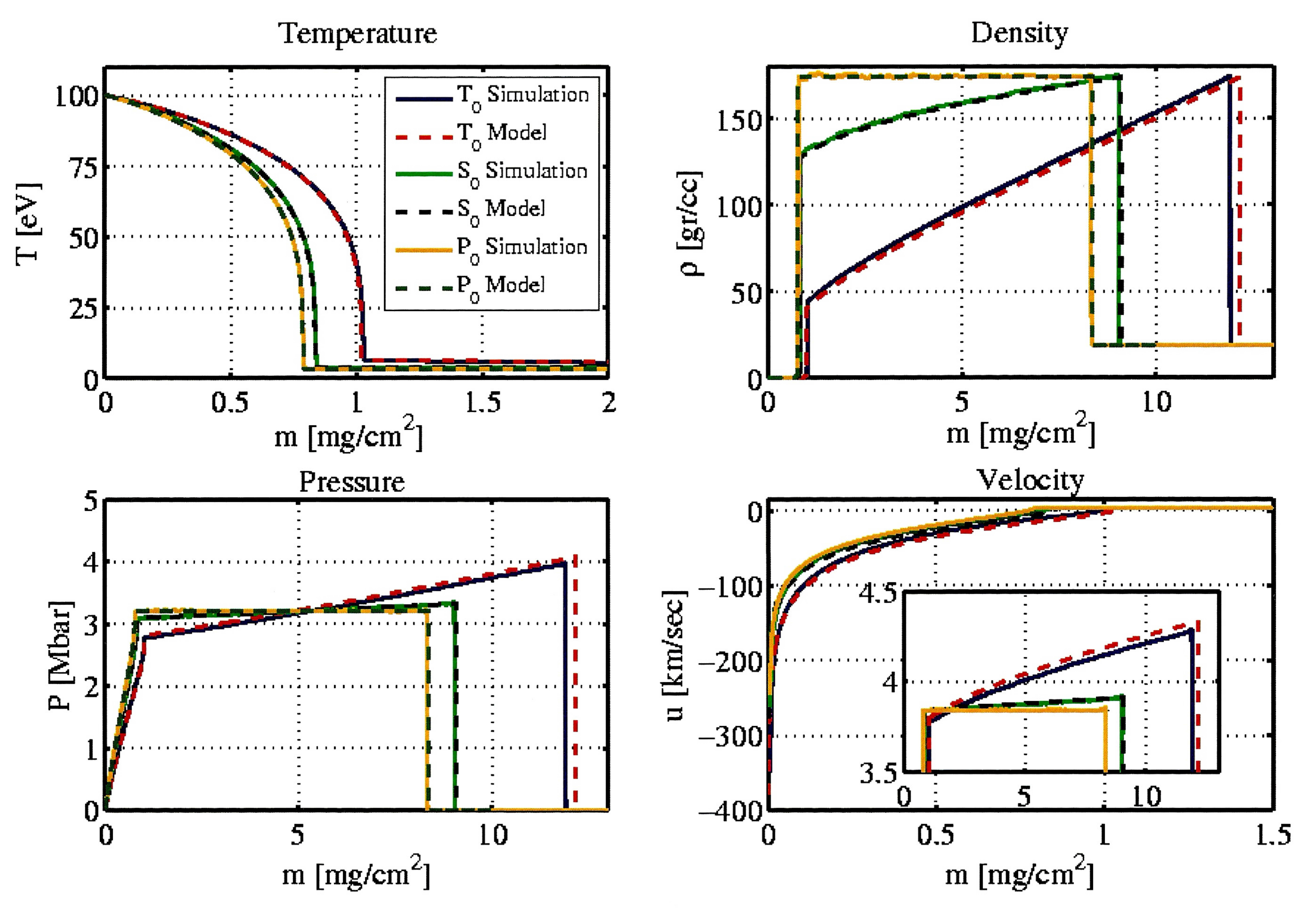}
}
\caption{(Color online) A comparison between the model (dashed lines) and simulations (full lines), under the different 
boundary conditions used in Figs. \ref{early_T0}-\ref{early_P0} in late times. Presented are the profiles of the temperature, density, pressure and matter velocity. The small box in the velocity
graph is a zoomed look on the shock region.}
\label{late}
\end{figure}

In Figs. \ref{early_T0}-\ref{late} the full analytic solution is compared to numerical simulation results in different boundary conditions and times. We present three 
representative cases: constant temperature, constant absorbed flux which yields $\uptau=0.123$, and constant ablation pressure which yields $\uptau=0.17$.
In Figs. \ref{early_T0}-\ref{early_P0} we present the comparison in relatively early times ($0.05\leqslant t \leqslant 0.15 \mathrm{nsec}$).
There is a good match between the analytic solution and the full numerical simulations,
as the results agree to within $1\%$ in the ablation region, and in the shock region, the agreement is about $5\%$.

The lack of perfect agreement between the full solution and the analytic one, may be due to the fact that in early times,
the heat front is not subsonic (and the ablation density is not infinite). The transition from supersonic to subsonic also invalidates
the approximation of negligible flow velocities, until the subsonic flow becomes established.

In Fig. \ref{late} we present a comparison in a later time, $t=1 \mathrm{nsec}$. The match between the model and the simulation is even better, as in the shock region they agree to within $2-3\%$. The better match at late times is
due to the build-up time of the solution. We note that the numeric solution front is $m_S$ and not at $m_F+m_S$
as could naively be expected. We assume that the shock front is weakly affected by the ablation front position, due to causal delay.

In addition, we see that the physical properties obtained using
the naive approximation of constant ablation pressure, significantly differ from the exact full solution (for example, see the density profile and shock front in Fig. \ref{late}).
This demonstrates the major benefits of using the model presented in this work, for evaluating the physical properties of the shock region.

\section{Discussion}
\label{discussion}

In this paper we investigated the diffusive radiative heat equation. At first we generalized the self-similar solutions of both
the supersonic and the subsonic heat equations, for a general boundary condition of the type $T=T_0t^{\uptau}$, and for power law opacity and heat capacity.
We used the technique shown in~\cite{ps} (that solved the equations for several specific cases) with the notation of~\cite{hr} (i.e. Super-transition-arrays based Au opacity). The numerical
results of $m_0$ and $e_0$ for Au are given in Fig. \ref{plots_super} for the supersonic case and in Fig. \ref{plots_sub} for the subsonic case, and can be used for general purposes, such as
hohlraum temperature evaluation in ICF~\cite{rosen,rosenScale,rosenScale2,rosenScale3,kauf}. 

Next, we focused on finding a self-similar solution of the shock region, under the assumption of a given pressure boundary condition of the type $P=P_0t^{\uptau_S}$ on
one boundary of the system, and the strong shock Hugoniot relations on the other. The solution was found to be in agreement with numerical simulation (Fig. \ref{Shock_check}).

Finally, We set the resulting $P_0$ and
$\uptau_S$ that the ablation heat wave solution yielded, substitute them as input to the shock solution, and patched both solutions at the interface. Using this method, we obtained a full analytic solution
to the non self-similar problem,
composed of two different solution, each one of them is self-similar by itself and is valid in a different region of the problem. This simple technique was tested with numerical
simulations and was found to be accurate to within $3\%$.

As mentioned in the Introduction, the shock wave can be evaluated via a naive approximation of constant ablation pressure~\cite{tlusty}, which is of course valid only if the ablation pressure
is relatively constant with time. Shock waves experiments have been analyzed using this approximation~\cite{kauf,kauf2}. In future work we plan to reproduce
the ablative shock analysis, and check
the agreement between this new model and experimental results.

\appendix
\section{A shooting method for solving the self-similar equations}
\label{nispach}

Obtaining a solution for the self-similar equations requires solving a set of equations with boundary conditions at both ends of the space. This is generally done using a shooting method, or using
the self-similarity of the equation.
In the case of the supersonic equation, Eq. \ref{super_ss}, one must find $\xi_F$ for which $\ti{T}(0)=1$. 
In both methods, $\xi_F$ is initially guessed. $\xi_F =1$ is usually
a good estimate. In the shooting method, each step the value of $\xi$ updates according to whether $\ti{T}(0)$ is larger
or smaller than one. The step size decreases with the number of iterations. A newton shooting method~\cite{ps} can easily
be applied for this problem. The self-similar method for finding $\xi_F$ is based on the fact that for a given time
and mass, the coordinate $\xi$ will only depend on the boundary temperature, according to Eq. \ref{super_dim_less}. This conservation of coordinates implies
\begin{equation}
\frac{\xi_1}{\xi_2}=(\frac{T_{0,1}}{T_{0,2}})^{\frac{\beta-\alpha-4}{2}}
\end{equation}
If $\ti{T}(0)\neq1$, $T_0$ in Eq. \ref{super_dim_less} must be multiplied by a normalization factor of $\ti{T}(0)$. Therefore the relation between the normalized $\xi_F$ and the guessed $\xi_g$ is
\begin{equation}
\frac{\xi_F}{\xi_g}=(\ti{T}(0))^{-\frac{\beta-\alpha-4}{2}}
\end{equation}

The subsonic set of equations can be solved in the same manner, using a double shooting method.
For a given value of $\xi_F$, initial values
for $\ti{P}(\xi_F),\p{\ti{P}}{\xi}|_{\xi_F}$ are guessed and the 
set of Eqs. \ref{sub_ss} is numerically integrated using a simple ODE solver. Then, the guess is updated according to whether $\ti{P}(0)$ is larger or smaller then $0$.
In the latter case, the integration diverges at some $\xi>0$. This procedure repeats for every iterative value $\xi_F$, while the determination of $\xi_F$ is done in a
manner similar to the supersonic solution, in order to normalize the self-similar quantities.

\begin{acknowledgments}
The authors are grateful to Yonatan Elbaz and Dov Shvarts for giving us the motivation to study this subject. The authors also wish to thank Ron Milo for reviewing the manuscript.
\end{acknowledgments}

\end{document}